\documentclass[preprintnumbers,amsmath,amssymb,floatfix,10pt,prd,onecolumn,superscriptaddress,nofootinbib]{revtex4}
\usepackage{epsfig,color}
\usepackage{epstopdf,float}
\usepackage{amssymb}
\usepackage{verbatim}
\usepackage{multirow}

\begin{document}

\title{\bf Thermodynamics in a Modified Gravity Involving Higher-Order Torsion Derivative Terms}

\author{Saira Waheed}
\email{swaheed@pmu.edu.sa}\affiliation{Prince Muhammad Bin Fahd
University, Khobar, Kingdom of Saudi Arabia}

\author{M. Zubair}
\email{mzubairkk@gmail.com;drmzubair@cuilahore.edu.pk}\affiliation{Department
of Mathematics, COMSATS University Islamabad, Lahore Campus,
Pakistan}

\begin{abstract}

The present study is elaborated to investigate the validity of
thermodynamical laws in a modified teleparallel gravity based on
higher-order derivatives terms of torsion scalar. For this purpose,
we consider spatially flat FRW model filled with perfect fluid
matter contents. Firstly, we explore the possibility of existence of
equilibrium as well as non-equilibrium picture of thermodynamics in
this extended version of teleprallel gravity. Here, we present the
first law and the generalized second law of thermodynamics (GSLT) using
Hubble horizon. It is found that non-equilibrium description of
thermodynamics exists in this theory with the presence of an extra
term called as entropy production term. We also establish GSLT using the
logarithmic corrected entropy. Further, by taking the equilibrium
picture, we discuss validity of GSLT at Hubble horizon for two $F$ different models. Using
Gibbs law and the assumption that temperature of matter within
Hubble horizon is similar to itself, we use different cases for choices
scale factors to discuss the GSLT validity graphically in all scenarios. It is found
that the GSLT is satisfied for a specified range of free
parameters in all cases.\\


\end{abstract}

\maketitle

\date{\today}

\section{Introduction}

Astronomical probes of modern cosmology suggest a speedy expanding
state of cosmos caused by a leading ingredient of obscure nature
present in the matter contents of cosmos, labeled as dark energy
(DE) \cite{1}. Numerous attempts have been made by the researchers
to investigate this mysterious component successfully. A list of
proposed candidates for DE is available in literature which is based
on one of the two different strategies namely modified matter source
models \cite{2} and modified gravitational theories \cite{2*}. On the basis of
their applications to various cosmological issues, a detailed
analysis of these candidates favors the modified gravitational
theories as the most successful tool for discussing different stages
of cosmic evolution. Some promising modified gravitational theories
include Gauss-Bonnet theory and its expended versions \cite{3},
$f(R)$ theory \cite{4} and its different generalizations involving
minimal or non-minimal interactions between different fields
(higher-order curvature correction terms, matter and scalar fields
as well as torsion scalar) like $f(R,T)$ \cite{5} and $f(R,T,Q)$ theories
\cite{5*}, scalar-tensor theories and its generalized versions
\cite{6} and the well-known teleparallel gravity with its different
extensions \cite{7}.

Teleparallel gravity is regarded as one of the interesting
alternative to Einstein's gravity (GR) in which torsional
formulation provides the gravitational source instead of curvature
scalar structure of GR \cite{8}. This theory is labeled as TEGR
(teleparallel equivalent of general relativity) and is determined by
the Lagrangian density involving curvature less Weitzenb$\ddot{o}$ck
connection instead of torsion less Levi-Civita connection along with
the vierbein as a fundamental tool. A variety of extended versions
of this theory have been presented in literature like $f(T)$ gravity
where a generic function of torsion scalar replaces the simple
torsion scalar term in the Lagrangian density \cite{7}. In this
respect, another different version of this theory has been proposed
by Kofinas and Saridakis \cite{9} where they introduced a new term
$T_G$ called teleparallel equivalent to Gauss-Bonnet term and then
further, they extended this theory to a more general case named as
$f(T, T_G)$ theory. Another significant modification is considered
by Harko et al. \cite{10} by including a non-minimal interaction of
torsion scalar with matter field in the action. In this respect,
another recent significant modification is $f(T,B)$ gravity
\cite{11}, where the term $B$ is related to the divergence of
torsion tensor and is termed as boundary term. This theory has been
tested by applying on different cosmological issues and found to be
very interesting \cite{12}. Another extended version of teleparallel
gravity has been proposed in literature \cite{13} which is based on
higher order derivative terms like $\nabla T$ and $\Box T$. The
basic motivation for the inclusion of such terms emerges from
already proposed other generalized versions of $f(R)$ gravity where
different higher order terms of Ricci scalars like
$R_{\mu\nu}R^{\mu\nu},~R_{\mu\nu\alpha\beta}R^{\mu\nu\alpha\beta},~(\nabla
R)^2$ etc., or its interaction with scalar as well as matter fields
are introduced in order to incorporate the quantum corrections
\cite{14}. Another idea behind it to formulate a fundamental gravity
like string theory where such terms arise in lagrangian density or
Kaluza Klein theories for reducing dimensions or to work at scales
close to plank scales in effective quantum gravity.

The description of thermodynamical picture of accelerating cosmos is
regraded as one of the most interesting issues in today's cosmology.
``The concepts of gravitation and thermodynamics are interlinked
with each other" is a fundamental connection supported by some
well-known results of thermodynamical study of black holes (BH). The
BH thermodynamics suggests the BHs as thermodynamical systems where
the terms like temperature and entropy are associated with the
geometrical quantities such as surface gravity and horizon area,
respectively \cite{15}. In this respect, the first effort was made
by Jacobson who used $T_hd\hat{S}_h=\delta\hat{Q}$ (Clausius
relation) along with $S=\frac{A}{4G}$ to derive the GR field
equations by taking Rindler model into account
($\hat{Q},~\hat{S},~T$ are notations for energy flux, entropy and
temperature, respectively) \cite{16}. Gibbons and Hawking \cite{17}
also made an attempt to explore these fundamental characteristics of
thermodynamics using de Sitter model. Frolov and Kofman \cite{18}
used flat quasi de-Sitter inflationary model of cosmos for
investigating such a connection of gravity and thermodynamics. They
concluded that the dynamical equations of Einstein gravity for
Friedmann model can be formulated using $dE=TdS$ with a slowly
rolling scalar field. It is seen that the Einstein field equations
for FRW universe can be obtained from the first law of
thermodynamics at the apparent horizon by making the use of
relationships for Hawking temperature and entropy given by
$T_A=\frac{1}{2\pi\tilde{R}_A}$ and $S_A=\frac{A}{4G}$,
respectively, where $A$ denotes the horizon area. Later on, this
connection was verified by Padmanabhan \cite{19} for a general
spherically symmetric spacetime. He found that the dynamical
equations for the considered model can be expressed in the form
$dE+PdV=TdS$. The question about the validity of such connection has
been already investigated in various contexts like braneworld
\cite{20}, Gauss-Bonnet gravity \cite{21}, the Lovelock gravity
\cite{22}, $f(R)$ gravity \cite{23,23*} and scalar-tensor theory
\cite{24}.

Karami and Abdolmaleki \cite{25} discussed the validity of GSLT in
$f(T)$ gravity using Hubble horizon and two viable models of $f(T)$
involving future singularities. They concluded that for present and
early eras, the GSLT remains valid while for later eras, it will be
satisfied for a specific value of torsion scalar. For $f(R,T)$ and
$f(R, T, R_{\mu\nu}T^{\mu\nu})$ theories, the study of
thermodynamics has been carried out by Sharif and Zubair \cite{26}
where they checked its validity at apparent horizon in
non-equilibrium perspective and they also formulated some possible
constraints on the coupling parameter. For a general Gauss-Bonnet
theory namely $f(G)$ gravity, Abdolmaleki and Najafi \cite{27} used
matter and radiation filled FRW geometry along with two different
$f(G)$ models to examine the validity of GSLT at dynamical apparent
horizon. Further this study has also been extended to the case of
$f(R,G)$ theory \cite{28}. The study of thermodynamical laws has
been also presented by Bahamonde et al. \cite{12} in a new modified
teleparallel theory which relates both $f(R)$ and $f(T)$ gravities
by the equation $R=-T+B$, where $B$ is the boundary term. They found
that this theory suggests the existence of non-equilibrium
thermodynamics picture due to the presence of additional entropy
production term. Further, by including the coupling of scalar field
with torsion and boundary term, the validity of GSLT has been
investigated at apparent horizon with and without including
logarithmic corrected entropy relation \cite{29}.

In a recent paper \cite{31}, GSLT validity has been explored by
Azizi and Borhani in a teleparallel gravity involving a non-minimal
coupling of torsion and matter and obtained interesting results. The
validity of GSLT has also been explored in $f(T,T_G)$ theory and the
possible constraints on the coupling parameter in terms of recent
cosmic parameters and power law solution \cite{32}. Sharif and
Waheed \cite{33} checked the validity of GSLT at Hubble, apparent,
particle and event horizons in a scalar-tensor gravity involving
chameleonic field as well as magnetic field effects. They concluded
that the GSLT valid in all cases for small red shift values. In
another study \cite{34}, the same authors investigated its validity
in Brans-Dicke theory by introducing power law and logarithmic
corrected entropy relations.

In the present paper, we will focus on the validity of GSLT at
Hubble horizon in both equilibrium and non-equilibrium perspectives
using a higher-order torsion derivatives based modified gravity. In
the coming section, we will present some basic notions of this
theory and the assumptions used for this work. Section \textbf{III}
formulates the possible forms of first as well as GSL of
thermodynamics and discuss the existence of its resulting
non-equilibrium picture. For this purpose, we will consider two
viable forms of $F$ function and some interesting cases of scale
factor. We also investigate its validity using logarithmic corrected
entropy there. In section \textbf{IV}, we investigate the existence
of equilibrium thermodynamics picture and check the validity of GSLT
using same cases of function $F$ as well as the scale factor. Last
section will summarize the whole discussion by highlighting the
major results.

\section{Basic Formulation of $F(T,(\nabla T)^2, \Box T)$ Gravitational Theory}

In this section, we will briefly present some basic formulation of
the modified teleparallel theory under consideration. Here we will
also specify the respective field equations along with the
assumptions taken for this work. The relation of metric and vierbein
$e^\mu_A$, the dynamical field of teleparallel gravity, is given by
\begin{equation}\label{1n}
g_{\mu\nu}=\eta_{AB}e^{A}_\mu e^{B}_\nu.
\end{equation}
The torsion tensor describing the gravitational field in terms of
Weitzenb\"{o}ck connection ($\Gamma^{\lambda}_{\nu\mu}\equiv
e^\lambda_A \partial_\mu e^A_\nu$) is expressed as
\begin{equation}\label{2n}
T^\rho_{\mu\nu}=e^\rho_A(\partial_\mu e^A_\nu-\partial_\nu e^A_\mu).
\end{equation}
The Lagrangian densities of teleparallel theory and its modified
versions are based on the torsion scalar obtained by the
contractions of the torsion tensor (\ref{2n}) as follows
\begin{equation}\label{3n}
T\equiv
\frac{1}{4}T^{\rho\mu\nu}T_{\rho\mu\nu}+\frac{1}{2}T^{\rho\mu\nu}T_{\nu\mu\rho}-{T_{\rho\mu}}^\rho
T^{\nu\mu}_\nu.
\end{equation}
The generalization of torsion based theories obtained by including
higher-order derivative terms like $(\nabla T)^2$ and $\Box T$ can
be expressed by the following action \cite{13}:
\begin{equation}\label{4n}
\mathcal{A}=\frac{1}{2{\kappa}^2}\int{dx^4e F(T, (\nabla T)^2, \Box
T)+ S_m(e^A_\rho, \psi_m)},
\end{equation}
where $S_m(e^A_\rho, \psi_m)$ denotes the ordinary matter part of
action. Here $\kappa^2=8\pi{G}$ and $F$ is a generic function of
torsion scalar and its higher-order derivatives. Also,
$e=det(e^A_\mu)=\sqrt{-g}$. Further, these higher-order derivatives
can be calculated by the formulas as
\begin{eqnarray}\label{5n}
(\nabla T)^2&=&\eta^{AB}e^\mu_A e^\nu_B \nabla_\mu T\nabla_\nu T=
g^{\mu\nu}\nabla_\mu T\nabla_\nu T, \\\label{6n} \Box
T&=&\eta^{AB}e^\mu_A e^\nu_B \nabla_\mu\nabla_\nu
T=g^{\mu\nu}\nabla_\mu\nabla_\nu T.
\end{eqnarray}
For the sake of simplicity in calculations, we introduce the
notations for higher-order derivatives as: $X_1=(\nabla T)^2$ and
$X_2=\Box T$. It is worthwhile to mention here that the action of
simple $f(T)$ gravity can be recovered by removing the higher-order
derivative terms, i.e., $X_1=X_2=0$. In terms of these new
notations, the respective field equations obtained by the variation
of the action (\ref{4n}) with respect to vierbein can be written as
\begin{eqnarray}\nonumber
&&\frac{1}{e}\partial_\mu(eF_T {e_{A}}^\tau
{S_{\tau}}^{\rho\mu})-F_T{e_{A}}^{\tau}
{S_{\nu}}^{\mu\rho}{T^{\nu}}_{\mu\tau}+\frac{1}{4}{e_{A}}^{\rho}F+
\frac{1}{4}\sum^2_{i=1}\{F_{X_{i}}\frac{\partial X_i}{\partial
{e^{A}}_\rho}\\\nonumber&&-\frac{1}{e}\left[\partial_\mu\left(eF_{{X}_{i}}\frac{\partial
X_i}{\partial\partial_\mu
{e^{A}}_\rho}\right)-\partial_\mu\partial_\nu\left(eF_{{X}_{i}}\frac{\partial
X_i}{\partial\partial_\mu\partial_\nu
{e^{A}}_\rho}\right)\right]\}\\\label{7n}&&
-\frac{1}{4e}\partial_\lambda\partial_\mu\partial_\nu\left(eF_{X_{2}}\frac{\partial
X_2}{\partial_\lambda\partial_\mu\partial_\nu
{e^{A}}_\rho}\right)=\frac{1}{2}{e_{A}}^{\tau}{{\mathcal{T}^{(m)}}_{\tau}}^\rho.
\end{eqnarray}
Here we have used the term ``superpotential" expressed in terms of
contortion tensor ${K^{\mu\nu}}_\rho\equiv
-\frac{1}{2}({T^{\mu\nu}}_\rho-{T^{\nu\mu}}_\rho-{T_{\rho}}^{\mu\nu})$
and is defined by the following relations:
$${S_{\rho}}^{\mu\nu}\equiv
\frac{1}{2}({K^{\mu\nu}}_\rho+{\delta^\mu}_\rho
T^{\theta\nu}_{\theta}-\delta^\nu_\rho T^{\theta\mu}_\theta).$$
Further, the notations $F_T$ and ${F_{X}}_i;~(i=1,2)$ stand for the
derivatives of the generic function $F$ with respect to the
subscript variable, i.e., $\frac{\partial F}{\partial
T},~\frac{\partial F}{\partial X_{i}}$, respectively. Also, the
contribution of ordinary matter given on left side of (\ref{7n}) can
be defined as follows
$${e_{A}}^{\tau}{\mathcal{T}^{(m)}}_\tau^\rho\equiv -\frac{1}{e}\frac{\delta S_m}{\delta{e^A}_\rho}.$$

Consider the spatially flat FRW universe geometry with cosmic radius
$a(t)$ given by the line element
\begin{equation}\label{8n}
ds^2=dt^2-a^2(t)(dx^2+dy^2+dz^2).
\end{equation}
The corresponding set of vierbein components are
\begin{equation}\nonumber
e^{A}_{\mu}=diag(1,~a(t),~a(t),~a(t))
\end{equation}
Here the energy-momentum tensor of ordinary matter source is assumed
to be perfect fluid given by
$$T_{{\mu}{\nu}}=({\rho}_m+p_m)u_{\mu}u_{\nu}-p_mg_{{\mu}{\nu}},$$
where $\rho_m$ and $p_m$ represent the density and pressure of
ordinary matter, respectively. Under these assumptions, the field
equations finally take the following form:
\begin{eqnarray}\nonumber
&&F_TH^2+(24H^2F_{X_1}+F_{X_2})(3H\dot{H}+\ddot{H})H+F_{X_2}\dot{H}^2+(3H^2-\dot{H})H\dot{F}_{X_2}\\\label{9n}
&&+24H^3\dot{H}\dot{F}_{X_1}+H^2\ddot{F}_{X_2}+\frac{F}{12}=\frac{\rho_m}{6},\\\nonumber
&&F_T\dot{H}+H\dot{F}_T+24H[2H\ddot{H}+3(\dot{H}+H^2)\dot{H}]\dot{F}_{X_1}+12H\dot{H}\dot{F}_{X_2}
+24H^2\dot{H}\ddot{F}_{X_1}\\\nonumber
&&+(\dot{H}+3H^2)\ddot{F}_{X_2}+24H^2F_{X_1}\dddot{H}+H\dddot{F}_{X_2}+24F_{X_1}\dot{H}^2(12H^2+\dot{H})\\\label{10n}
&&+24HF_{X_1}(4\dot{H}+3H^2)\ddot{H}=-\frac{p_m}{2},
\end{eqnarray}
where $H=\dot{a}/a$ represents the Hubble parameter and the dot
denotes the cosmic time rate of change. Equations (\ref{9n}) and
(\ref{10n}) can be rearranged to following forms:
\begin{eqnarray}\label{11n}
3H^2={\kappa^2}_{eff}\rho_{eff},\quad
\dot{H}=-{\kappa^2}_{eff}(\rho_{eff}+p_{eff}),
\end{eqnarray}
where the effective energy density and pressure are the combinations
$\rho_{eff}=\rho_m+\rho_{T}$ and $p_{eff}=p_m+p_T$, respectively.
Also, the effective coupling is defined as
${\kappa^2}_{eff}=\frac{\kappa^2}{2F_T}$. These contributions of
density and pressure due to torsion are given by
\begin{eqnarray}\nonumber
\rho_{T}&=&\frac{1}{\kappa^2}[-6(24H^2F_{X_1}+F_{X_2})(3H\dot{H}+\ddot{H})H)-6F_{X_2}\dot{H}^2
-6(3H^2-\dot{H})H\dot{F}_{X_2}\\\label{12n}
&-&144H^3\dot{H}\dot{F}_{X_1}-6H^2\ddot{F}_{X_2}-\frac{F}{2}],\\\nonumber
p_{T}&=&\frac{2}{\kappa^2}[H\dot{F}_T+24H[2H\ddot{H}+3(\dot{H}+H^2)\dot{H}]\dot{F}_{X_1}+12H\dot{H}\dot{F}_{X_2}
+24H^2\dot{H}\ddot{F}_{X_1}\\\nonumber
&+&(\dot{H}+3H^2)\ddot{F}_{X_2}+24H^2F_{X_1}\dddot{H}+H\dddot{F}_{X_2}+24F_{X_1}\dot{H}^2(12H^2+\dot{H})\\\label{13n}
&+&24HF_{X_1}(4\dot{H}+3H^2)\ddot{H}-3H^2F_T].
\end{eqnarray}
For this spatially flat geometry, the torsion scalar and its
derivatives $(\nabla T)^2$ and $\Box T$ turn out to be
\begin{equation}\label{14n}
T=-6H^2, \quad X_1=144H^2\dot{H}^2, \quad
X_2=-12\left[\dot{H}(\dot{H}+3H^2)+H\ddot{H}\right].
\end{equation}
Also, the ordinary matter satisfies the usual continuity equation
and is given by
\begin{equation}\label{15n}
\dot{\rho}_m+3H(\rho_m+p_m)=0.
\end{equation}
It is worthwhile to mention here that similarly, the effective
density and pressure satisfies the continuity equation
\begin{equation}\nonumber
\dot{\rho}_{eff}+3H(\rho_{eff}+p_{eff})=0
\end{equation}
which consequently gives rise to the non-conservation of its torsion
scalar counterparts due to the presence of an extra term on left
side as follows
\begin{equation}\label{16n}
\dot{\rho}_T+3H(\rho_T+p_T)=\frac{T}{\kappa^2}\dot{F}_T.
\end{equation}
By assuming the barotropic equation of state
$p_m=\omega_m\rho_m;~0\leq\omega_m\leq1$, the integration of the
continuity equation leads to the following relation
\begin{equation}\nonumber
\rho_m=\rho_{m0}a^{-3(1+\omega_m)},
\end{equation}
where $\rho_{m0}$ represents an arbitrary constant of integration.

\section{Non-Equilibrium and Equilibrium Perspectives of Thermodynamics in $F(T, X_1, X_2)$ Gravity}

In this section, we present a brief discussion on the first and
generalized second law of thermodynamics by considering the
perspective of non-equilibrium. It has already been discussed in
literature \cite{26,35,36} that such picture exits in the extended
gravitational theories based on curvature or torsion matter
couplings like $f(R,T),~f(R,T,Q),~f(T,L_m)$ and $f(R,L_m)$ theories.

\subsection{First Law of Thermodynamics}

Here we describe the possible form of first thermodynamical law in
this modified gravity and investigate the issue of non-equilibrium
picture there. For a flat FRW geometry, the radius of dynamical
apparent horizon in terms of $h^{\alpha\beta}$ given by the
condition
$h^{\alpha\beta}\partial_{\alpha}\tilde{R}_A\partial_{\beta}\tilde{r}_A=0$,
takes the form
\begin{equation}\label{12n}
\tilde{R}_A=\frac{1}{H}.
\end{equation}
Its time rate of change yields the following equation:
\begin{equation}\nonumber
\frac{d\tilde{R}_A}{dt}=\tilde{R}^{3}H\kappa^2_{eff}(\rho_{eff}+p_{eff}).
\end{equation}
After simplifying, the above equation can be written as
\begin{equation}\label{}
\frac{F_Td\tilde{R}_A}{G}=4\pi\tilde{R}_A^3H(\rho_{eff}+p_{eff})dt.
\end{equation}
The area of the horizon is defined as $A=4\pi\tilde{R}_A^2$ and the
temperature associated with this horizon in terms of surface gravity
$\kappa_{sg}$ is defined by $T_A=\kappa_{sg}/2\pi$, where
\begin{eqnarray}\nonumber
\kappa_{sg}=\frac{1}{2\sqrt{-h}}\partial_{\alpha}(\sqrt{-h}h^{\alpha\beta}\partial_\beta
\tilde{R}_A).
\end{eqnarray}
For flat FRW model, it will take the form
\begin{equation}
-\frac{1}{\tilde{R}_A}\left(1-\frac{\dot{\tilde{R}}_A}{2H\tilde{R}_A}\right)=-\frac{\tilde{R}_A}{2}(2H^2+\dot{H}).
\end{equation}
On multiplication by the factor
$\left(1-\frac{\dot{\tilde{R}}_A}{2H\tilde{R}_A}\right)=-2\pi\tilde{R}T_A$,
the above equation leads to the following relation:
\begin{eqnarray}\nonumber
&&T_Ad\left(\frac{AF_T}{4G}\right)=-(4\pi\tilde{R}_A^3Hdt-2\pi\tilde{R}_A^2\dot{\tilde{R}}_A)
(\rho_{eff}+p_{eff})+\frac{\pi\tilde{R}^2_A}{G}T_AdF_T.
\end{eqnarray}
The Bekenstein-Hawking entropy relation \cite{15} suggests $S=A/4G$.
Like many other modified gravity theories (for example, \cite{26,
33, 35, 36}), this relation is modified by the inclusion of
$G_{eff}$ instead of $G$. Consequently, in this theory, it takes the
form $S_A=\frac{AF_T}{4G}$. Thus the last equation can be re-written
as
\begin{equation}\label{13n}
T_Ad\tilde{S}_A=(2\pi\tilde{R}_A^2\dot{\tilde{R}}_A-4\pi\tilde{R}_A^3Hdt)(\rho_{eff}+p_{eff})+\frac{\pi\tilde{R}^2_A}{G}T_AdF_T.
\end{equation}
The Misner-Sharp energy defined by the relation
$E=\frac{\tilde{R}_A}{2G_{eff}}$ or equivalently, $E=\rho_{eff}V$
provides the total matter energy density of universe (a sphere of
radius $\tilde{R}_A$ at the apparent horizon). Here the volume of
the universe is given by the equation $V=4/3\pi\tilde{R}_A^3$. In
this modified teleparallel gravity, this relation leads to
\begin{eqnarray}\nonumber
dE&=&4\pi{\tilde{R}}_A^2(\rho_m+\rho_{T}){d}\tilde{R}_A-4\pi\tilde{R}_A^3H(\rho_{eff}+p_{eff})dt+\frac{{\tilde{R}^3}_A}{2G}(dF_T).
\end{eqnarray}
Inserting this $dE$ in Eq.(\ref{13n}), we obtain
\begin{equation}\label{14n}
T_Ad\tilde{S}_A=dE+2\pi\tilde{r}_A^2(p_{eff}-\rho_{eff})d\tilde{R}_A+\frac{\tilde{R}_A}{G}\left(3+\pi\tilde{R}_AT_A\right)dF_T.
\end{equation}
Also, the total work density is defined by the equation \cite{32}
\begin{equation}\label{15n}
W=-\frac{1}{2}\left(T^{(m)\alpha\beta}h_{\alpha\beta}+\tilde{T}^{(de)\alpha\beta}h_{\alpha\beta}\right)=\frac{1}{2}
(\rho_{eff}-p_{eff}).
\end{equation}
Where the notations $T^{(m)\alpha\beta}$ and $\tilde{T}^{(de)}$
stand for the energy densities due to ordinary and dark matter,
respectively. Introducing work density in Eq.(\ref{14n}) leads to
the final form of first law of thermodynamics given by
\begin{equation}\label{16n}
T_Ad\tilde{S}_A+T_Ad\tilde{S}_p=dE-WdV,
\end{equation}
where the term
$d\tilde{S}_p=\frac{\tilde{R}_A}{GT_A}\left(3+\pi\tilde{R}_AT_A\right)dF_T$
is due to the entropy production term in non-equilibrium
thermodynamics. Thus, we conclude that in this extended teleparallel
gravity, the form of first law of thermodynamics is modified by the
presence of a surplus term. This in agreement with the already
available results in literature for $f(R)$, $f(R,T)$, $f(R,T,Q)$
theories as well as generalized Gauss-Bonnet gravity where a surplus
term exist giving rise to non-equilibrium thermodynamics there.

\subsection{GSLT in Modified $f(T)$ Gravity}

In the present section, we explore the issue of GSLT validity in the
context of this generalized teleparallel gravity. The GSLT suggests
that function obtained by the sum of entropies of horizon and
ordinary matter fluid components always increases versus cosmic
time. This issue has been already investigated in the context of
various modified theories like $f(R)$, $f(R)$ theory involving
matter geometry coupling, $f(T),~f(R,T),~f(R,T,Q),~f(R,L_m)$ and
scalar-tensor theories. Here we will utilize new form of first law
of thermodynamics obtained in the previous section. Mathematically,
GSLT can be written as
\begin{equation}\label{17n}
\dot{\tilde{S}}_{total}=\dot{\tilde{S}}_h+\dot{\tilde{S}}_{p}+\dot{\tilde{S}}_{in}\geq0,
\end{equation}
where the notations $\tilde{S}_h,~\tilde{S}_p$ and $\tilde{S}_{in}$
stand for horizon entropy, entropy production term and entropy of
matter components inside horizon, respectively. First law of
thermodynamics (\ref{16n}) provides the relation:
\begin{equation}\nonumber
T_id\tilde{S}_i=dE_i+p_idV-T_id\tilde{S}_p
\end{equation}
which can also be written as
\begin{equation}\nonumber
T_{in}\dot{\tilde{S}}_{in}=(\rho_i+p_i)4\pi{\tilde{R}^2}_A\left(\dot{\tilde{R}}_A-H\tilde{R}_A\right)
+\frac{4}{3}\pi\tilde{R}^3_AQ_i-T_{in}\dot{\tilde{S}}_p,
\end{equation}
where $T_{in}$ denotes the temperature for all components inside the
horizon, $Q_i$ represents the ith term interaction component. Taking
summation of all inside horizon components entropies, we get
$$\sum Q_i=0, ~~ \sum(\rho_i+p_i)=\rho_{eff}+p_{eff}.$$
Consequently, we have
\begin{equation}\nonumber
T_{in}\dot{\tilde{S}}_{in}=(\rho_{eff}+p_{eff})4\pi{\tilde{R}^2}_A\left(\dot{\tilde{R}}_A-H\tilde{R}_A\right)-T_{in}\dot{\tilde{S}}_p.
\end{equation}
Further, after an easy calculation, one can write the last equation
as follows:
\begin{equation}\label{18n}
\dot{\tilde{S}}_{in}+\dot{\tilde{S}}_{p}=\frac{4\pi}{G}\frac{\dot{H}(\dot{H}+H^2)F_T}{(2H^2+\dot{H})H^3}.
\end{equation}
Also, from the Bekenstein-Hawking entropy relation, one can find
\begin{equation}\label{19n}
\dot{\tilde{S}}_h=\frac{\pi}{GH^2}\left(\dot{F}_T-2\frac{\dot{H}}{H}F_T\right).
\end{equation}
Thus, from Eqs.(\ref{17n}), (\ref{18n}) and (\ref{19n}), the GSLT
constraint takes the following form
\begin{eqnarray}\label{20n}
\dot{\tilde{S}}_{tot}=\frac{4\pi}{G}\left(\frac{\dot{H}(\dot{H}+H^2)F_T}{(2H^2+\dot{H})H^3}
+\frac{1}{4H^2}\{\dot{F}_T-2\frac{\dot{H}}{H}F_T\}\right)\geq0.
\end{eqnarray}
In the upcoming subsections, we will explore the validity of this
constraint using two different functional forms of $F$ and in last,
by considering the logarithmic entropy correction term.

\subsubsection{The Validity of GSLT Constraint for a Function
Independent of $X_2$}

Here we will explore the validity of GSLT using the form of $F$ that
is independent of $X_2$ given as follows
\begin{equation}\label{1*}
F(T,~X_1,~X_2)=T+\frac{\alpha_1X_1}{T^2}+\alpha_2e^{\frac{\delta
X_1}{T^4}},
\end{equation}
where $\alpha_1,~\alpha_2$ and $\delta$ are all dimensionless
constants. This form of $F$ has already been used in literature for
checking the validity of energy constraints as well as the stability
using fixed point theory \cite{13,1*}. The GSLT constraint for this
functional form is given by
\begin{eqnarray}\nonumber
\dot{\tilde{S}}_{tot}&=&\frac{4\pi}{G}[\frac{\dot{H}^2+\dot{H}H^2}{H^3(2H^2+\dot{H})}
\left(1-\frac{2\alpha_1X_1}{T^3}-\frac{4\alpha_2\delta
X_1}{T^5}e^{\frac{\delta
X_1}{T^4}}\right)+\frac{1}{4H^2}[\{(\frac{6\alpha_1X_1}{T^4}\\\nonumber
&+&\left(\frac{16\alpha_2\delta^2X_1^2}{T^{10}}+\frac{20\alpha_2\delta_1}{T^6}\right)e^{\frac{\delta
X_1}{T^4}})\dot{T}+(-\frac{2\alpha_1}{T^3}-\frac{4\alpha_2\delta}{T^5}\left(1+\frac{\delta}{T^4}\right)e^{\frac{\delta
X_1}{T^4}})\dot{X}_1\}\\\label{20}
&-&\frac{2\dot{H}}{H}\{1-\frac{2\alpha_1X_1}{T^3}-\frac{4\alpha_2\delta
X_1}{T^5}e^{\frac{\delta X_1}{T^4}}\}]]\geq0.
\end{eqnarray}
Now we will discuss the validity of GSLT constraint (\ref{20}) by
taking four different expansions of scale factor given as follows
\begin{itemize}
\item Constant Hubble parameter: $H=H_0$, where $H_0$ is recent value of Hubble
parameter, i.e., the de Sitter model.
\item Expressing the higher order time rates in terms of cosmographic parameters like $q,~r,~s$ etc.
\item Power law form: $a(t)=a_0(t_s-t)^{-b}$, where $a_0$ is the
present value of the scale factor and $t_s\geq t,~b>0$.
\item Intermediate form: $a(t)=e^{b_1t^\beta}$, where $b_1$ is any
positive constant and $0<\beta<1$.
\end{itemize}
In the first place, we evaluate the GSLT for the choice of de-Sitter
model having constant Hubble parameter $H=H_0$. In this case, it is
found that GSLT is trivially satisfied as all the derivatives vanish
for this choice.

Secondly, we discuss the validity of GSLT constraint by introducing
the cosmographic parameters. Here we define some interesting
cosmographic parameters depending on higher-order derivatives of
scalae factor obtained by the Taylor's series expansion of scale
factor like deceleration, jerk, snap and lerk parameters etc.. These
parameters are defined as follows
\begin{eqnarray}\nonumber
q=-\frac{1}{H^2}\frac{a^{(2)}}{a}, \quad
j=\frac{1}{H^3}\frac{a^{(3)}}{a}, \quad
s=\frac{1}{H^4}\frac{a^{(4)}}{a}, \quad
l=\frac{1}{H^5}\frac{a^{(5)}}{a}.
\end{eqnarray}
It is worthwhile to mention here that all the higher-order
derivatives of Hubble parameter can be expressed as a linear
combination of these cosmographic parameters. For example, first
four order time rates of Hubble parameter in terms of these
cosmographic parameters can be written as
\begin{eqnarray}\nonumber
\dot{H}&=&-H^2(1+q),\quad \ddot{H}=H^3(j+3q+2),\quad
\dddot{H}=H^4(s-2j-5q-3),\\\label{i}
H^{(4)}&=&H^5\left(l-5s+10(q+2)j+30(q+2)q+24\right).
\end{eqnarray}
Consequently, the terms like torsion scalar, $X_1$ and $X_2$ and
their corresponding time rates can be expressed in terms of these
cosmographic parameters as given below
\begin{eqnarray}\nonumber
T&=&-6H^2, \quad X_1=144H^6(1+q)^2,\quad
X_2=-12\left(H^4(1+q)^2-3H^4(1+q)+H^4(j+3q+2)\right),\\\nonumber
\dot{T}&=&12H^3(1+q), \quad
\dot{X}_1=-288H^7(1+q)\{(1+q)^2+j+3q+2\},\\\label{i1}
\dot{X}_2&=&12H^5\{3(1+q)(j+3q+2)-3(j+3q+2)-6(1+q)^2(s-2j-5q-3)\}.
\end{eqnarray}
In this case, for the graphical analysis, we consider the present
values of these cosmographic quantities as suggested in literature
\cite{2*} and are given by
$H_0=0.718,~q_0=-0.64,~j_0=1.02,~s_0=-0.39$ and $l_0=4.05$. For
these values, the quantities of Eq.(\ref{i1}) become
\begin{eqnarray}\nonumber
T=-3.0931, \quad X_1=2.5569, \quad X_2=-0.4771, \quad
\dot{T}=1.5590, \quad \dot{X}_1=-12.5409, \quad \dot{X}_2=-0.8654.
\end{eqnarray}
By using these values, we explored the possible ranges of model
parameters namely $\alpha_1,~\alpha_2$ and $\delta$ using region
graph as given in Figure \textbf{1}. The detailed possible ranges of
these parameters for which GSLT condition remain valid are given in
Table \textbf{I}.

Now we consider the possibility of power law form of expansion
factor given by the relation $a(t)=a_0(t_s-t)^{-b},~b>0,~t_s\geq t$.
Here the point $t=t_s$ leads to the presence of a Big Rip
singularity in this super accelerated cosmos model \cite{3*}. In
this case, the Hubble parameter, Torsion scalar and the terms
$X_1,~X_2$ along with their derivative terms turn out to be as
follows
\begin{eqnarray}\nonumber
H&=& \frac{b}{t_s-t}, \quad T=-\frac{6b^2}{(t_s-t)^2}, \quad
\dot{H}=\frac{b}{(t_s-t)^2},\quad
\dot{T}=-\frac{12b^2}{(t_s-t)^3},\\\nonumber
X_1&=&\frac{144b^4}{(t_s-t)^6},\quad
\dot{X}_1=\frac{864b^4}{(t_s-t)^7}, \quad
X_2=-\frac{36b^2(1+b)}{(t_s-t)^4}, \quad
\dot{X}_2=-\frac{144b^2(1+b)}{(t_s-t)^5}.\\\label{i2}
\end{eqnarray}
Here $b$ is the power law parameter and $\alpha_1$, $\alpha_2$,
$\delta$ are the model parameters. In this discussion, we fix
$\delta$ and evaluate the validity ranges for $\alpha_1$ and
$\alpha_2$. If $\delta>0$ then validity of GSLT requires
$\alpha_2\geqslant0$ for all values of $\alpha_1$ whereas if
$\delta\leqslant0$ then GSLT is satisfied for $\alpha_1\geqslant0$
with all values of $\alpha_2$ as shown in Table \textbf{II}. The
graphical illustration of validity of GSLT constraint is shown in
Figure \ref{fig2} for some particular cases. In some cases it isn't
obvious to find the exact region of validity, one of such cases is
shown in right plot of Figure \ref{fig2}. In left of Figure
\ref{fig2}, we have selected one particular validity range and
presented its evolution of GSLT for different values of $\delta$.
\begin{figure}[H]
\epsfig{file=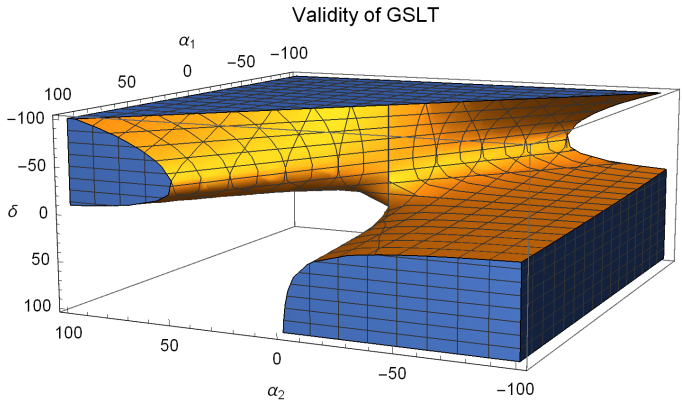,width=0.5\linewidth}\caption{The plot
represents the validity regions for GSLT constraint in terms of
cosmographic parameters for model (\ref{1*}).}\label{fig1}
\end{figure}
\begin{figure}[H]
\epsfig{file=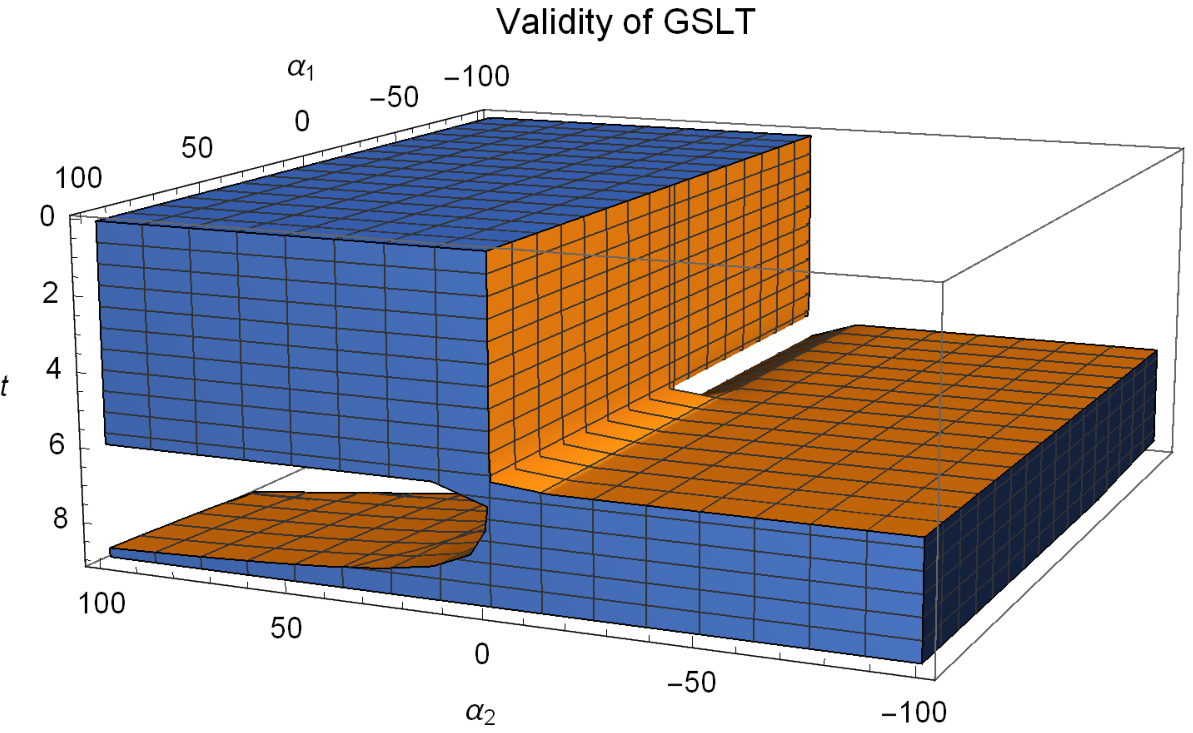,width=0.5\linewidth}\epsfig{file=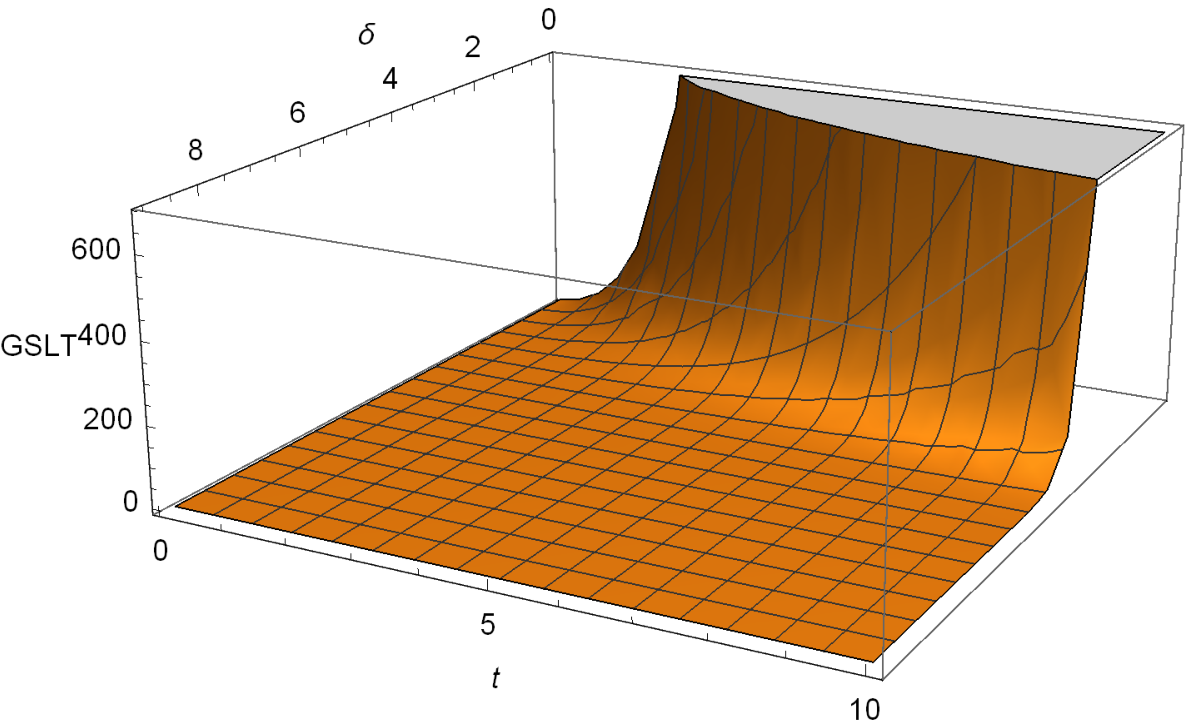,
width=0.5\linewidth} \caption{Left plot represents the regions where
GSLT is satisfied for $\delta=10$ and right graph corresponds to
evolution of GST versus $\delta$ for $\alpha_1=-0.002$,
$\alpha_2=0.001$, $b=2$ and $t_s=0.9$.}\label{fig2}
\end{figure}

Now we will discuss the validity of the GSLT constraint using the
intermediate form of expansion radius given by
$a(t)=e^{b_1t^\beta},~0<\beta<1,~b>0$. Such form of expansion factor
is very significant as it plays an important role in the description
of inflationary scenario and hence compatible with the astrophysical
evidences \cite{4*}. For this form of expansion radius, the
cosmological parameters like Hubble parameter, torsion scalar, terms
$X_1,~X_2$ and its first order time rates take the following form
\begin{eqnarray}\nonumber
H&=&b_1\beta t^{\beta-1},\quad
\dot{H}=b_1\beta(\beta-1)t^{\beta-2},\quad
T=-6b_1^2\beta^2t^{2(\beta-1)},\quad
\dot{T}=-12b_1^2\beta^2(\beta-1)t^{2\beta-3},\\\nonumber
X_1&=&144b_1^4\beta^4(\beta-1)^2t^{4\beta-6},\quad
\dot{X}_1=288b_1^4\beta^4(\beta-1)^2(2\beta-3)t^{4\beta-7},\\\nonumber
X_2&=&-12\{b_1^2\beta^2(\beta-1)(2\beta-3)t^{2\beta-4}+3b_1^3\beta^3(\beta-1)t^{3\beta-4}\},\\\label{i3}
\dot{X}_2&=&-12\{b_1^2\beta^2(\beta-1)(2\beta-3)(2\beta-4)t^{2\beta-5}
+3b_1^3\beta^3(\beta-1)(3\beta-4)t^{3\beta-5}\}.
\end{eqnarray}
In intermediate form, we have two additional parameters $b$ and
$\beta$ along with the model parameters $\alpha_1$, $\alpha_2$ and
$\delta$. Herein, we set $b_1=2$, $\sigma=2$ and $\beta=0.5$. We fix
the parameter $\delta$ to explore the validity of GSLT depending on
the parameters $\alpha_1$ and $\alpha_2$. If $\delta>0$,then
validity of GSLT requires $\alpha_1\geqslant0$ and
$\alpha_2\leqslant-20$ whereas if $\delta<0$ it requires
$\alpha_1\geqslant0$ and $\alpha_2\leqslant-5$. Moreover, in case of
$\delta=0$, one need to set $\alpha_1\leqslant0$ along with all values
$\alpha_2$.
\begin{figure}[H]
\epsfig{file=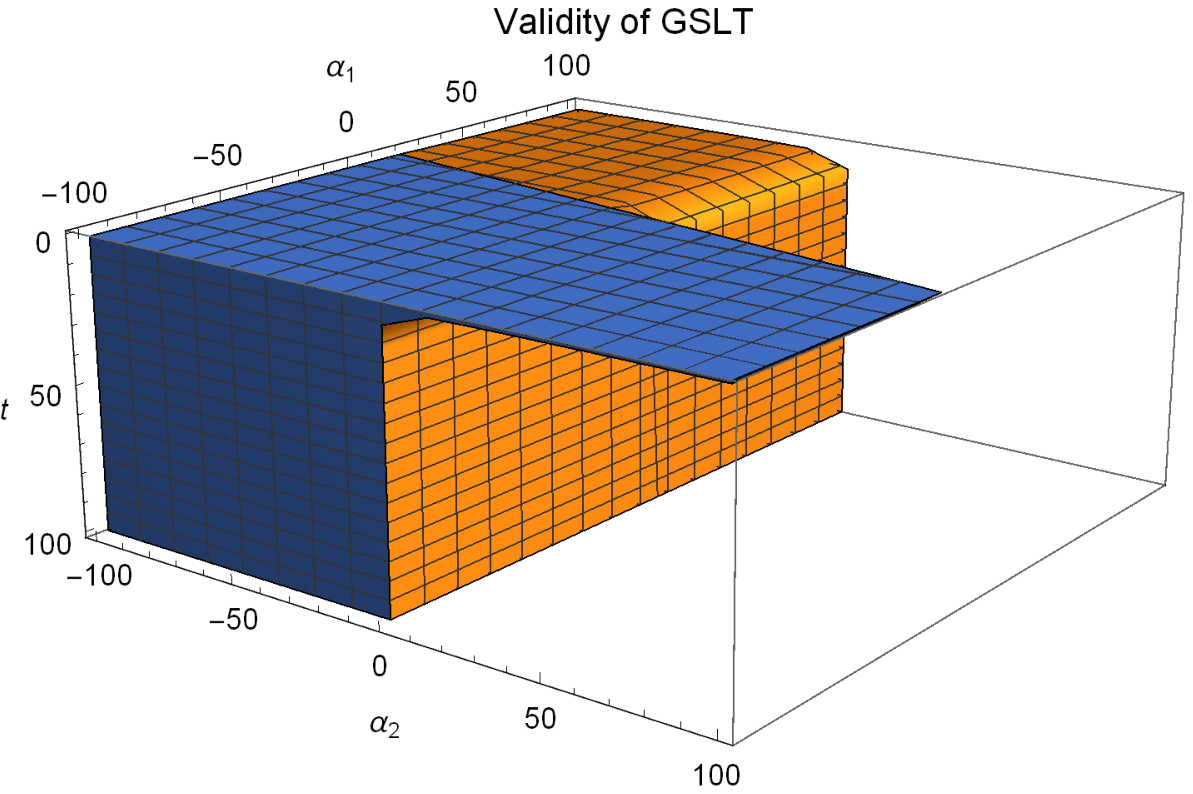,width=0.5\linewidth}\epsfig{file=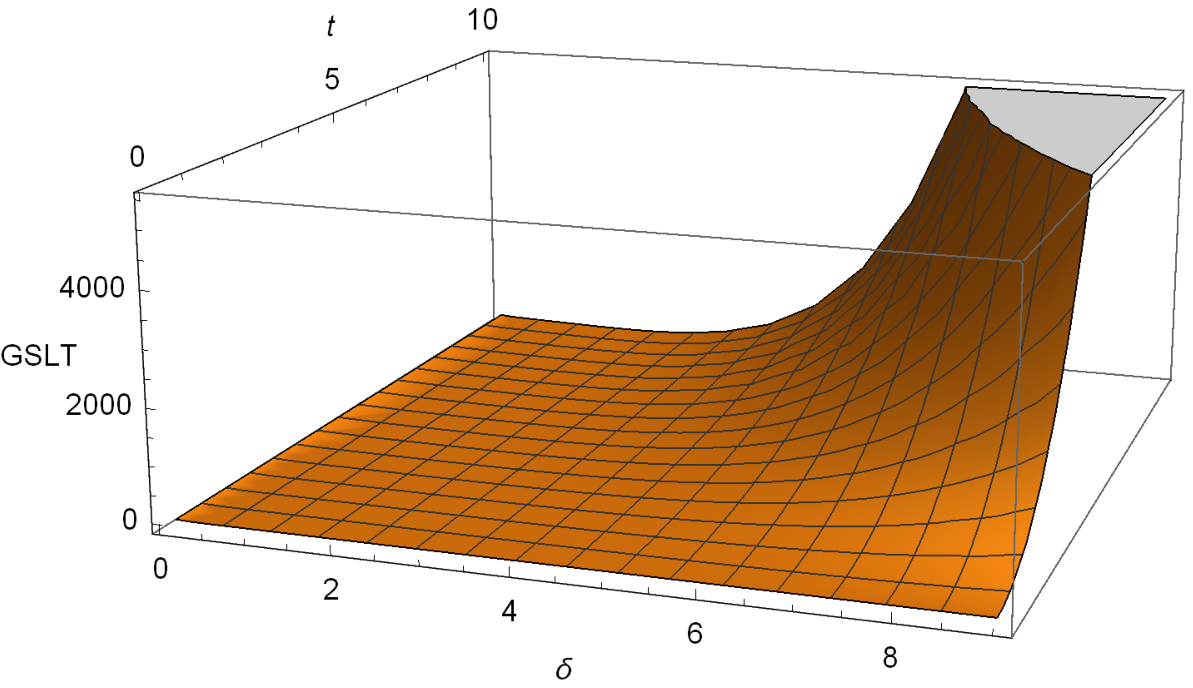,
width=0.5\linewidth} \caption{Left plot represents the regions where
GSLT is satisfied for $\delta=10$ in intermediate case and right
graph corresponds to evolution of GST versus $\delta$ for
$\alpha_1=0.2$ and $\alpha_2=-21$.}\label{fig3}
\end{figure}
In left plot of Figure \ref{fig3}, we represent one particular
region of validity whereas in the right plot, we use the validity
range of parameters $\alpha_1$ and $\alpha_2$ to show the evolution
of GSLT versus $\delta$.

\subsubsection{The Validity of GSLT Constraint for a Function
Independent of $X_1$}

Now we will consider the form of generic function $F(T, X_1, X_2)$
independent of the term $X_1$ which is defined by the following
relation
\begin{eqnarray}\label{1}
F(T, X_1, X_2)=T+\frac{\beta_1
X_2}{T}+\frac{\beta_2X_2^2}{T^3}+\beta_3e^{\frac{\sigma X_2}{T^3}},
\end{eqnarray}
where $\beta_i; ~i=1, 2, 3$ and $\sigma$ are all arbitrary constant
parameters. In this case, the GSLT constraint will take the form
\begin{eqnarray}\nonumber
\dot{\tilde{S}}_{tot}&=&\frac{4\pi}{G}[\{\frac{\dot{H}(\dot{H}+H^2)}{(2H^2+\dot{H})H^3}-\frac{\dot{H}}{2H^3}\}
\{1-\frac{\beta_1X_2}{T^2}-\frac{3\beta_2X_2^2}{T^4}-\frac{3\sigma\beta_3X_2}{T^4}e^{\frac{\sigma
X_2}{T^3}}\}\\\nonumber
&+&\frac{1}{4H^2}\{\left(\frac{2\beta_1X_2}{T^3}+\frac{12\beta_2X_2^2}{T^5}
+3\sigma\beta_3X_2\left(\frac{4}{T^5}+\frac{3\sigma
X_2}{T^4}\right)e^{\frac{\sigma X_2}{T^3}}\right)\dot{T}\\\label{2}
&+&\left(-\frac{\beta_1}{T^2}-\frac{6\beta_2X_2}{T^4}-\frac{3\sigma\beta_3}{T^4}
\left(1+\frac{\sigma}{T^3}\right)e^{\frac{\sigma
X_2}{T^3}}\right)\dot{X}_2\}]\geq0.
\end{eqnarray}
Using previously defined four different cases of expansion factor
namely constant Hubble parameter, cosmographic parameters, power law
and intermediate forms along with the corresponding relations of
torsion scalar, $X_1$ and $X_2$ with their time rates given by
Eqs.(\ref{i1}-\ref{i3}), we will check the compatibility of this
GSLT condition and explore the possible choices of free model
parameters. Here, we have four model parameters $\beta_1$,
$\beta_2$, $\beta_3$ and $\sigma$. If $\beta_2=0$, then one can
retrieve the results similar to the previous model. Herein, we set
$b=1$ and $\sigma=2$. For de Sitter case, it is seen that the GSLT
constraint trivially holds. For the case of cosmographic parameters,
by using the previously defined present values of these parameters,
the possible ranges of $\beta_1,~\beta_2$ and $\beta_3$ are explored
as shown in Figure \textbf{4} and listed in Table \textbf{I}.

In power law model, we fix the parameter $\beta_3$ and find the
values of other parameters $\beta_1$ and $\beta_2$. It is found that
GSLT validate for the choice of $\beta_3\leqslant0$ and the detailed
results are shown in Table \ref{Table2}. In Fig.\ref{fig5}, we
present graphical illustration of the validity range and show the
evolution of GSLT versus $\beta_3$ for the choice $\beta_1=.02$,
$\beta_2=0.3$ and $\sigma=2$.
\begin{figure}[H]
\epsfig{file=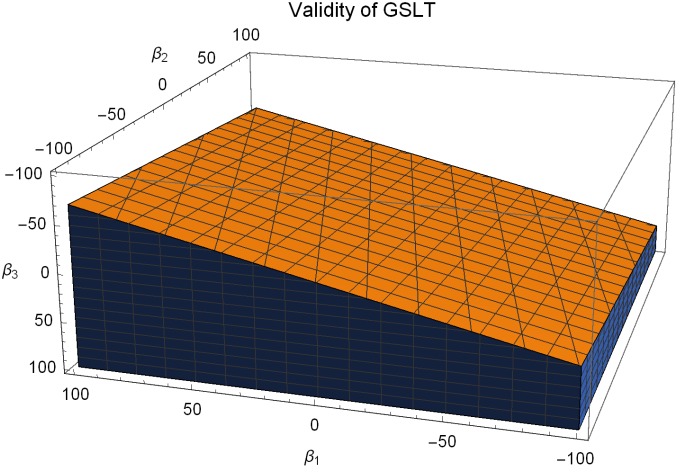, width=0.5\linewidth} \caption{The plot
represents the validity regions for GSLT constraint in terms of
cosmographic parameters for the model (\ref{1}).}\label{fig4}
\end{figure}
\begin{figure}[H]
\epsfig{file=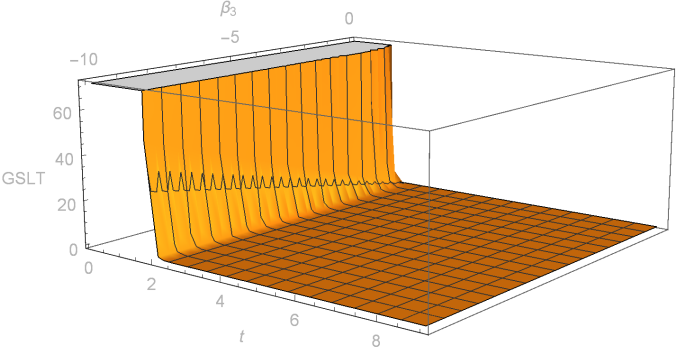,width=0.5\linewidth} \caption{The graphical
illustration of GSLT versus $\beta_3$ for the second model. Herein,
we set $\beta_1=.02$, $\beta_2=0.3$ and $\sigma=2$.}\label{fig5}.
\end{figure}

\begin{figure}[H]
\epsfig{file=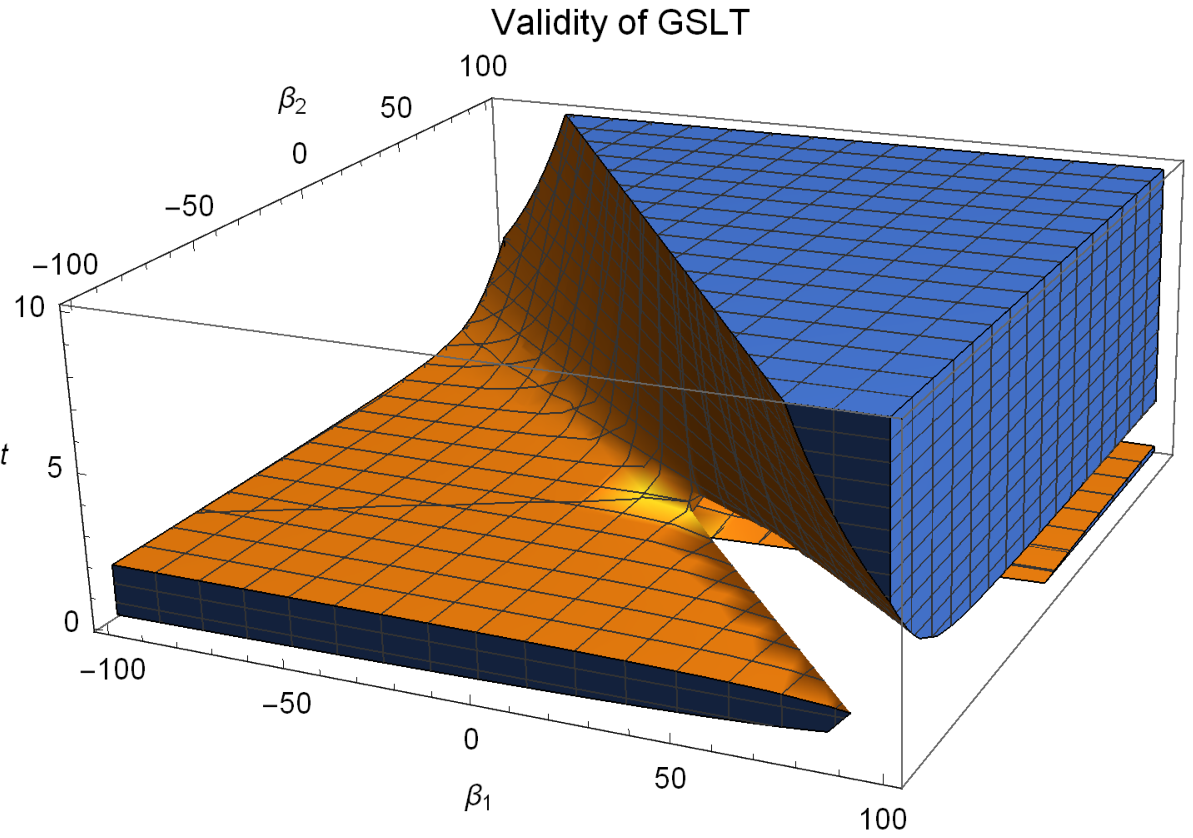,width=0.5\linewidth}\epsfig{file=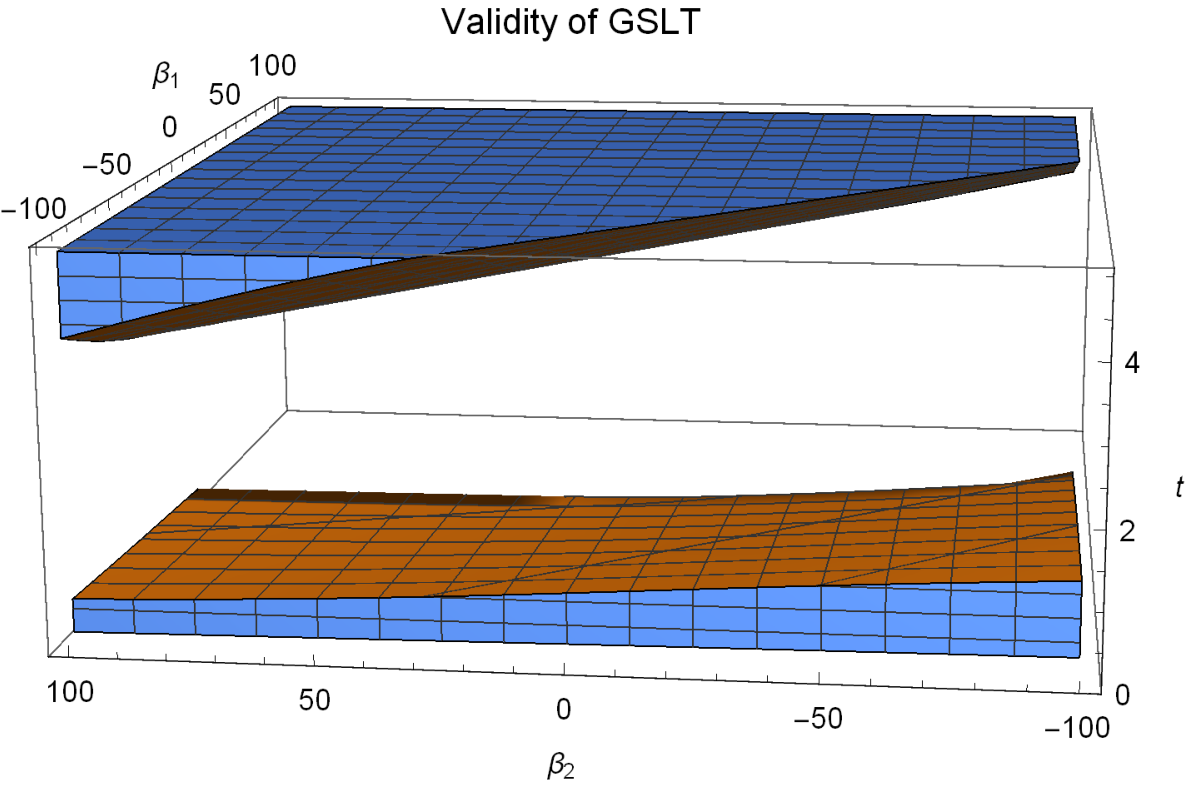,
width=0.5\linewidth} \caption{Left plot represents the regions where
GSLT is satisfied for $\beta_3=0$ in intermediate case and right
graph corresponds to evolution of GST versus $\beta_3=-2$. Herein,
we set $b_1=2$, $\sigma=2$ and $\beta=0.5$.}\label{fig6}
\end{figure}

In case of intermediate form, we have three parameters $\beta_1$,
$\beta_2$ and $\beta_3$. We fix one parameter $\beta_3$ to set the
validity ranges for the parameters $\beta_1$ and $\beta_2$ and
results are shown in Table \ref{Table2}. It is found that GSLT is
valid only for $\beta\geqslant0$, in case of $\beta\geqslant0$
validity region exists only for earlier times. In this discussion,
we find some cases where it is difficult to find valid regions. The
graphical illustration of some cases is shown in Figure \ref{fig6},
left plot shows the validity regions for $\beta_3=0$. It is found
that GSLT is not valid for the choice of $\beta<0$ except some
particular cases with specific regions. In the right plot of Figure
\ref{fig6}, we present the validity regions for $\beta_3=-2$.

\subsubsection{The Validity of GSLT Constraint for Logarithmic
Corrected Entropy}

\begin{figure}[H]
\epsfig{file=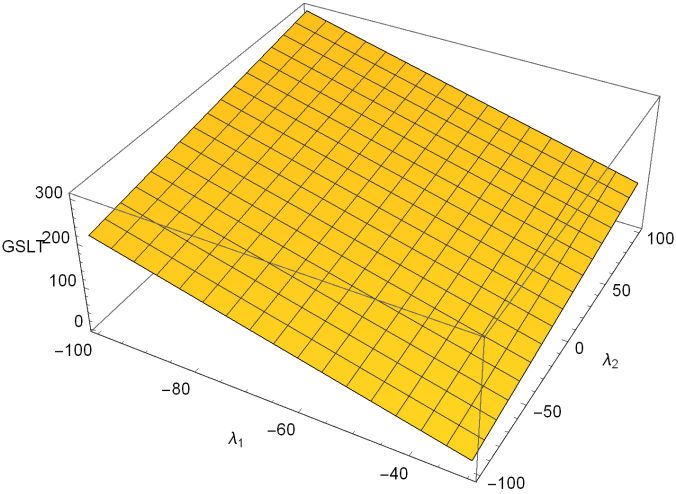,width=0.5\linewidth} \caption{Validity of GSLT
for logarithmic corrected entropy with constant Hubble
parameter.}\label{fig7}
\end{figure}

Here we will consider the entropy correction formula involving
logarithmic terms with effective gravitational coupling
$\kappa^2_{eff}$. In the present case, such entropy correction is
defined by the following equation \cite{37}
\begin{eqnarray}
\tilde{S}_{LCE}=\frac{AF_T}{4G}+\lambda_1\ln\left(\frac{AF_T}{4G}\right)+\lambda_2\left(\frac{4G}{AF_T}\right)+\lambda_3,
\end{eqnarray}
where $\lambda_i;~i=1,2,3$ are all non-zero dimensionless arbitrary
constants. In case of Hubble horizon, the time rate of this entropy
is given by
\begin{eqnarray}
\dot{\tilde{S}}_{LCE}=\frac{\pi}{GH^2}\left[1+\lambda_1\left(\frac{GH^2}{\pi
F_{T}}\right)-\lambda_2\left(\frac{GH^2}{\pi
F_T}\right)^2\right](\dot{F}_T-\frac{2\dot{H}}{H}F_T).
\end{eqnarray}
Consequently, the time rate of total entropy will become
\begin{eqnarray}\nonumber
\dot{\tilde{S}}_{tot}=\frac{4\pi}{G}\frac{\dot{H}(\dot{H}+H^2)F_T}{(2H^2+\dot{H})H^3}+
\frac{\pi}{GH^2}\left[1+\lambda_1\left(\frac{GH^2}{\pi
F_{T}}\right)-\lambda_2\left(\frac{GH^2}{\pi
F_T}\right)^2\right](\dot{F}_T-\frac{2\dot{H}}{H}F_T)\geq0.
\end{eqnarray}
For the functional form defined by the Eq.(\ref{1*}), this
constraint will take the form
\begin{eqnarray}\nonumber
\dot{\tilde{S}}_{tot}&=&\{\frac{4\pi}{G}\frac{\dot{H}(\dot{H}+H^2)}{(2H^2+\dot{H})H^3}-\frac{2\pi\dot{H}}{GH^3}
\left(1+\lambda_1\left(\frac{GH^2}{\pi
F_T}\right)-\lambda_2\left(\frac{GH^2}{\pi
F_T}\right)^2\right)\}\\\nonumber
&\times&\{1-\frac{2\alpha_1X_1}{T^3}-\frac{4\alpha_2\delta
X_1}{T^5}e^{\frac{\delta
X_1}{T^4}}\}+\frac{\pi}{GH^2}\left(1+\lambda_1\left(\frac{GH^2}{\pi
F_T}\right)-\lambda_2\left(\frac{GH^2}{\pi
F_T}\right)^2\right)\\\nonumber
&\times&\{(\frac{6\alpha_1X_1}{T^4}+\left(\frac{16\alpha_2\delta^2X_1^2}{T^{10}}+\frac{20\alpha_2\delta
X_1}{T^6}\right)e^{\frac{\delta
X_1}{T^4}})\dot{T}+(-\frac{2\alpha_1}{T^3}-\frac{4\alpha_2\delta}{T^5}e^{\frac{\delta
X_1}{T^4}}\\\label{3}
&\times&\left(1+\frac{\delta}{T^4}\right))\dot{X}_1\}\geq0.
\end{eqnarray}
For the function $F$ form given by the Eq.(\ref{1}), this GSLT
inequality can be written as
\begin{eqnarray}
\dot{\tilde{S}}_{tot}&=&\{\frac{4\pi}{G}\frac{\dot{H}(\dot{H}+H^2)}{(2H^2+\dot{H})H^3}-\frac{2\pi\dot{H}}{GH^3}
\left(1+\lambda_1\left(\frac{GH^2}{\pi
F_T}\right)-\lambda_2\left(\frac{GH^2}{\pi
F_T}\right)^2\right)\}\\\nonumber
&\times&\{1-\frac{\beta_1X_2}{T^2}-\frac{3\beta_2X_2^2}{T^4}-\frac{3\sigma
X_2\beta_3}{T^4}e^{\frac{\sigma
X_2}{T^3}}\}+\frac{\pi}{GH^2}\left(1+\lambda_1\left(\frac{GH^2}{\pi
F_T}\right)-\lambda_2\left(\frac{GH^2}{\pi
F_T}\right)^2\right)\\\nonumber
&\times&\{\left(\frac{2\beta_1X_2}{T^3}+\frac{12\beta_2X_2^2}{T^5}+3\sigma\beta_3X_2\left(\frac{4}{T^5}+\frac{3\sigma
X_2}{T^4}\right)e^{\frac{\sigma X_2}{T^3}}\right)\dot{T}\\\label{4}
&+&\left(-\frac{\beta_1}{T^2}-\frac{6\beta_2X_2}{T^4}-\frac{3\sigma\beta_3}{T^4}e^{\frac{\sigma
X_2}{T^3}}\left(1+\frac{\sigma}{T^3}\right)\right)\dot{X}_2\}\geq0.
\end{eqnarray}
Using four cases for expansion factor and the corresponding terms,
we will check the validity of these GSLT constraints given by
Eqs.(\ref{3}) and (\ref{4}). In case of constant Hubble parameter
$H_0$, the GSLT is satisfied if
$-16.8-2.78\lambda_1+0.46\lambda_2\geqslant0$. Here we find a
relation for the validity of GSLT which depends on two dimensionless
parameters $\lambda_1$ and $\lambda_2$. We show the evolution of
GSLT in Figure \ref{fig7}. For the cosmographic parameters, the GSLT
validity regions are explored in Figure \textbf{8}. Here the left
and right plots correspond to the GSLT constraints (\ref{3}) and
(\ref{4}), respectively. The detail possible ranges of free model
parameters $\alpha_1,~\alpha_2,~\delta,~\beta_1,~\beta_2$ and
$\beta_3$ for which GSLT conditions remain valid are listed in Table
\textbf{I}.

Next we search for validity regions in case power law model and the
results are presented depending on different values of $\delta$.
Some validity regions are presented in Figure \ref{fig9} for both
$\delta=2$ and $\delta=-2$ respectively. In case of intermediate
form of scale factor, we show the validity regions of GSLT for two
cases $\delta=0$ and $\delta>0$. Left plot of Figure \ref{fig10}
shows some particular validity regions for $\delta=0$ whereas in
right plot we select $\delta>0$. The results of validity regions are
shown in Table \ref{Table2}.
\begin{figure}[H]
\epsfig{file=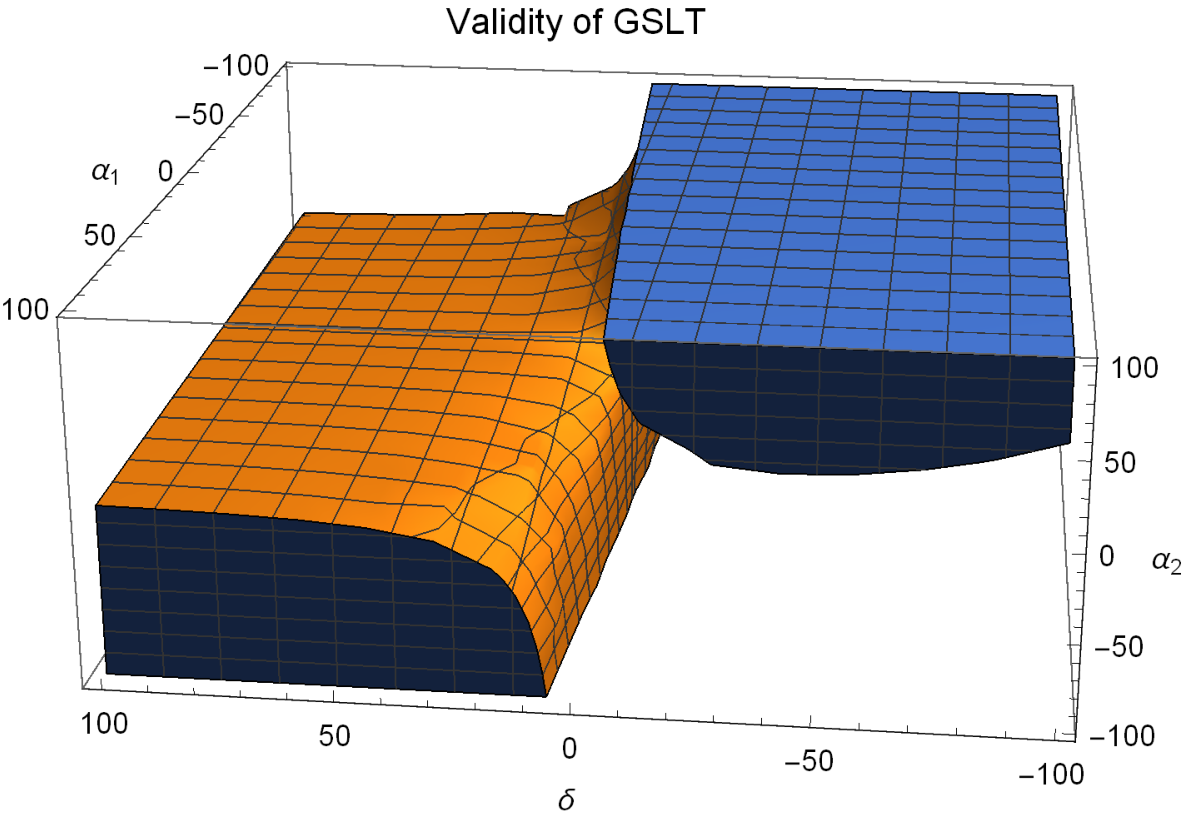,width=0.5\linewidth}\epsfig{file=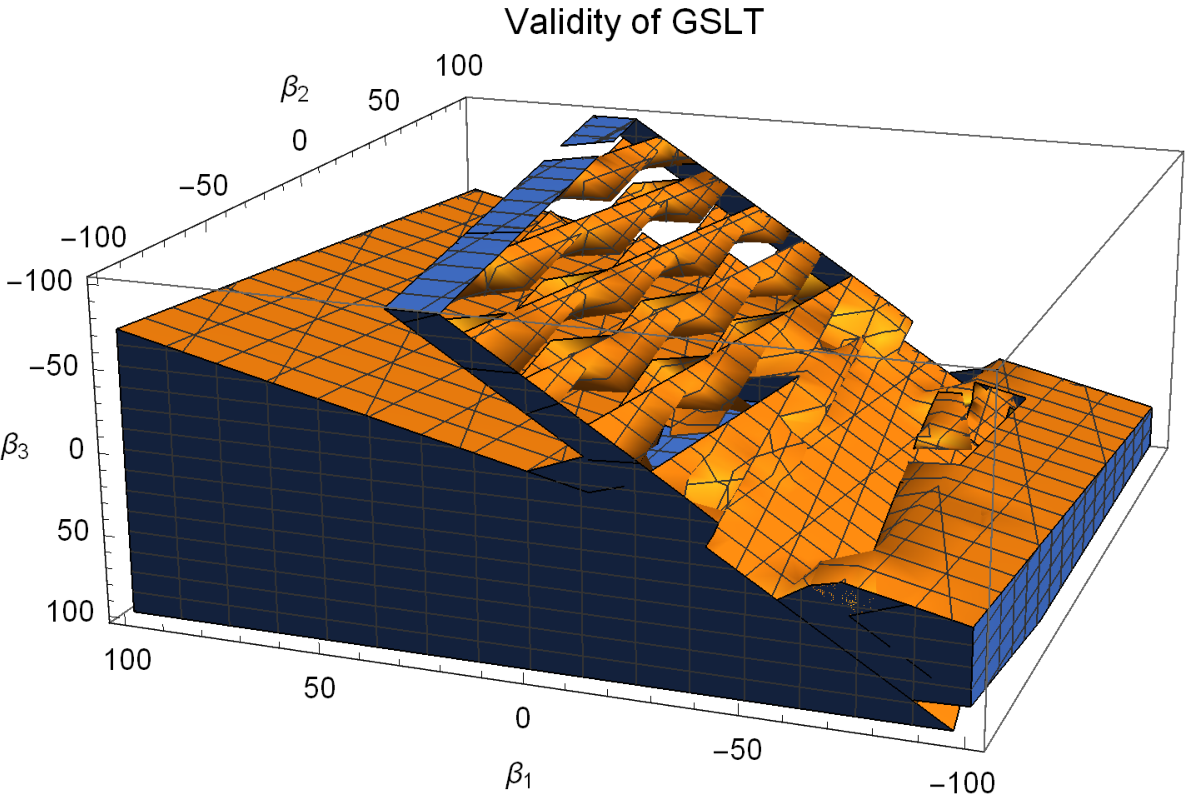,
width=0.5\linewidth} \caption{Left and right plots represent the
possible validity regions for logarithmic corrected entropy in terms
of cosmographic parameters for both $F$ models.}\label{fig8}
\end{figure}
\begin{figure}[H]
\epsfig{file=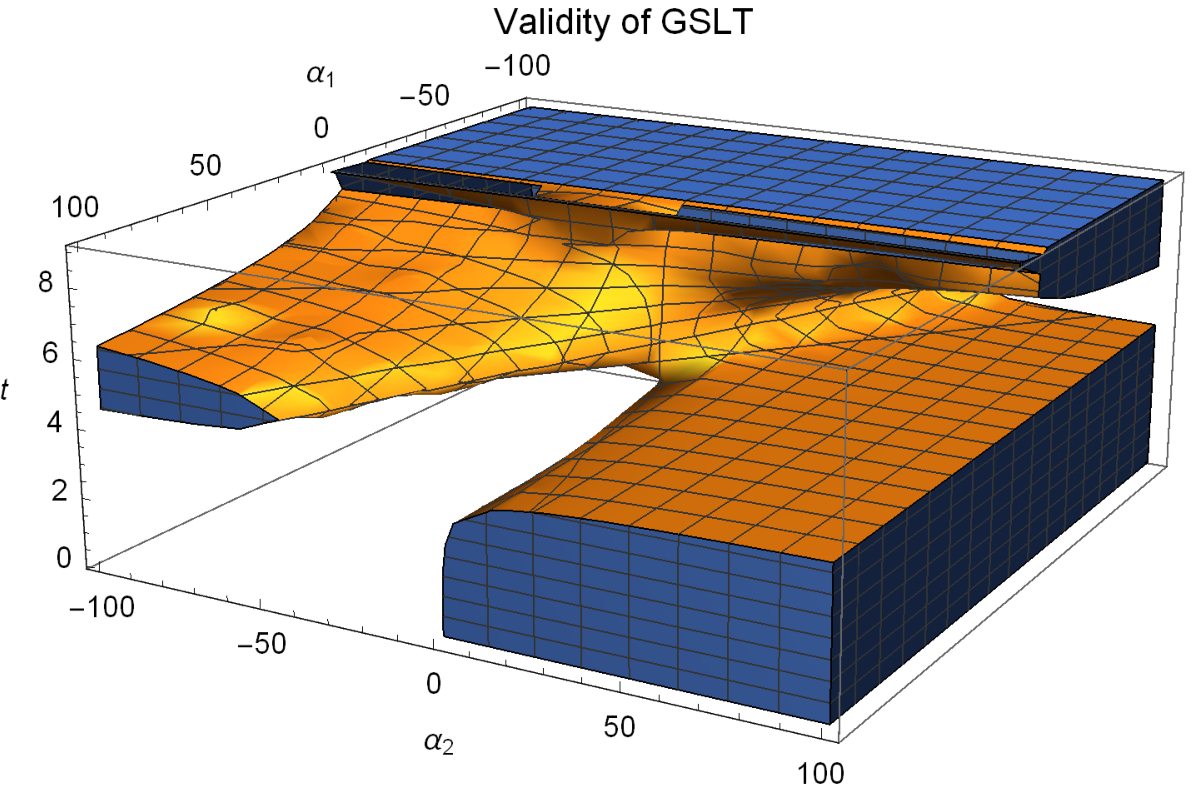,width=0.5\linewidth}\epsfig{file=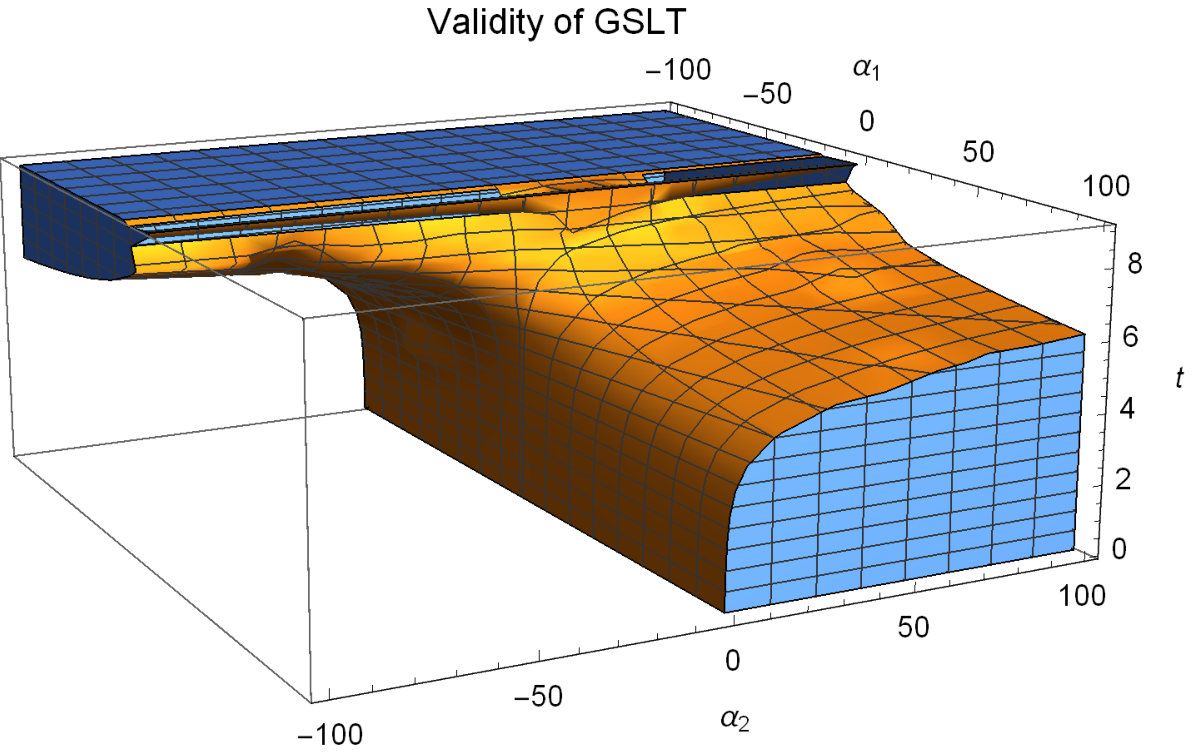,
width=0.5\linewidth} \caption{Left plot represents the validity
regions for logarithmic corrected entropy in power law case with
$\delta=2$, right graph corresponds to evolution of GST for
$\delta=-2$. Herein, we set $\lambda_1=2$ and
$\lambda_2=3$.}\label{fig9}
\end{figure}

We also examine the validity of GSLT constraint for logarithmic
corrected entropy in the background of function independent of
$X_1$. It is found that GSLT is trivially satisfied for de Sitter
model. In this case, we fix $\beta_3$ and develop the validity
regions depending on the values of $\beta_1$ and $\beta_2$ for both
power law and intermediate cases. In intermediate form particular
validity regions are shown in Figure \ref{fig11} and other validity
ranges can be seen in Table \ref{Table2}.
\begin{figure}[H]
\epsfig{file=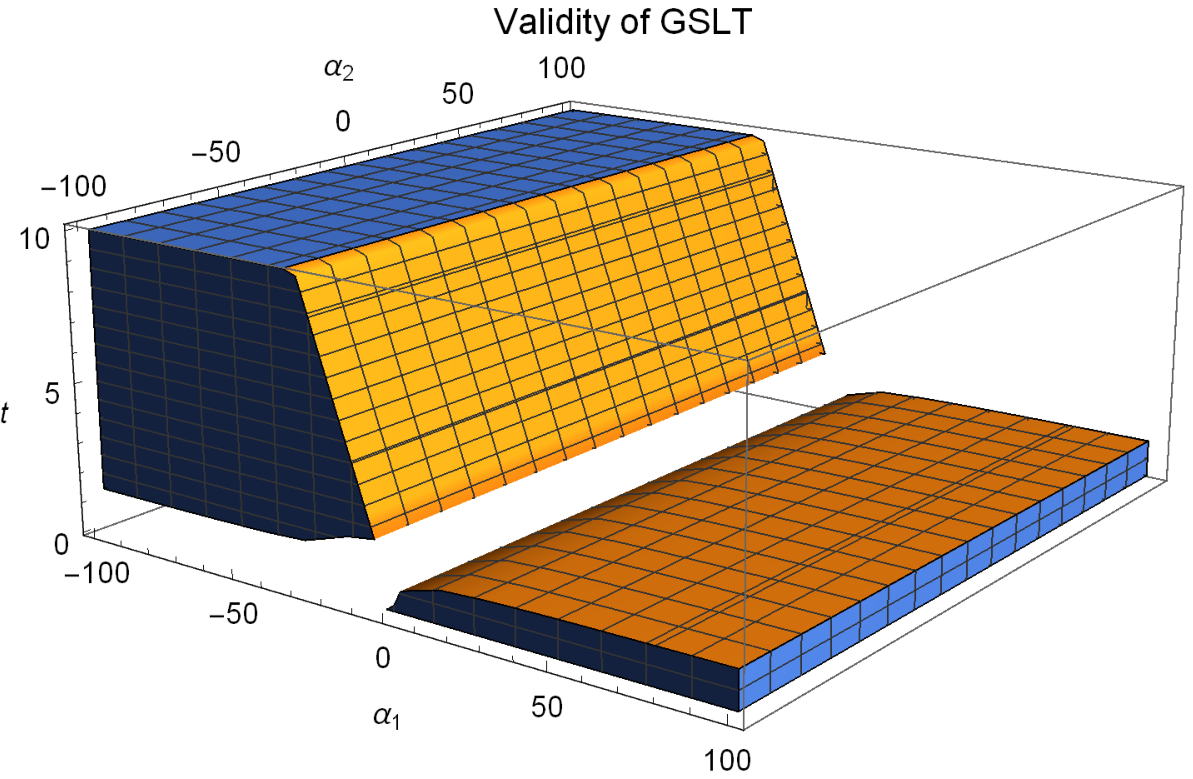,width=0.5\linewidth}\epsfig{file=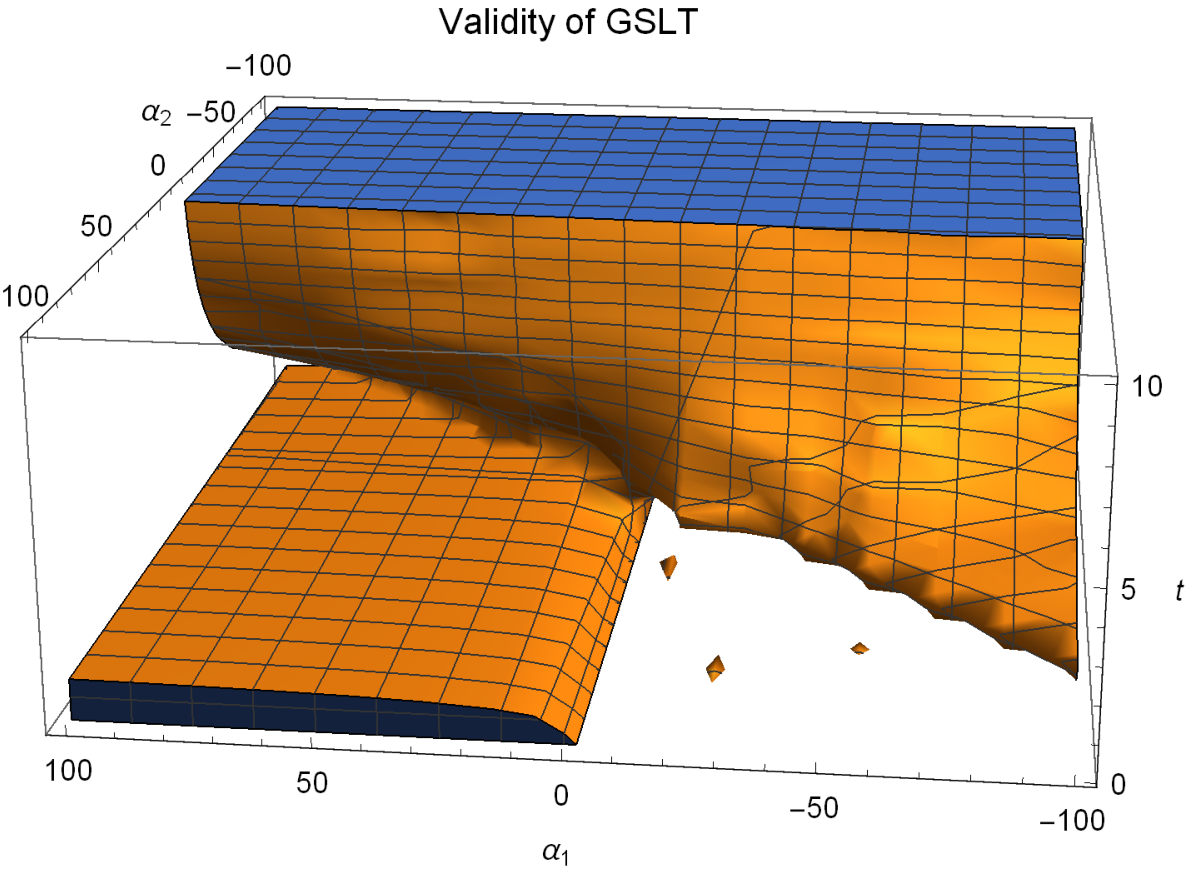,
width=0.5\linewidth} \caption{Left plot represents the validity
regions for logarithmic corrected entropy in intermediate case with
$\delta=0$, right graph corresponds to evolution of GST for
$\delta=2$. Herein, we set $\lambda_1=2$ and
$\lambda_2=3$.}\label{fig10}
\end{figure}
\begin{figure}[H]
\epsfig{file=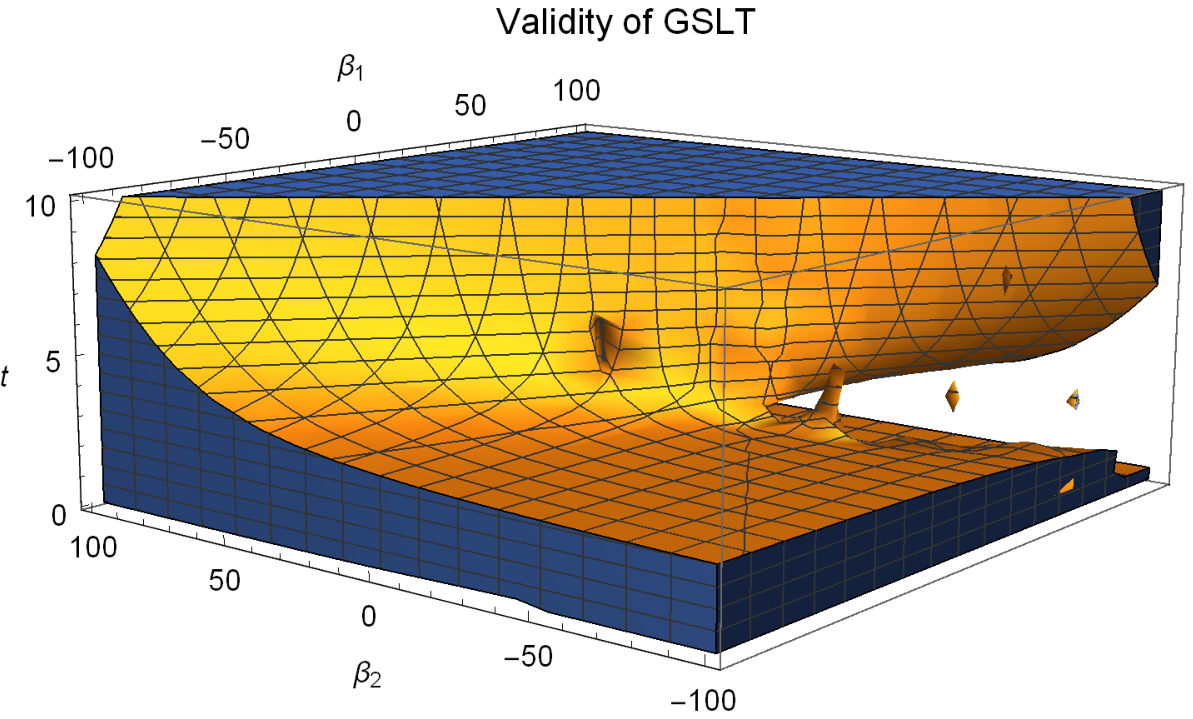,width=0.5\linewidth}\epsfig{file=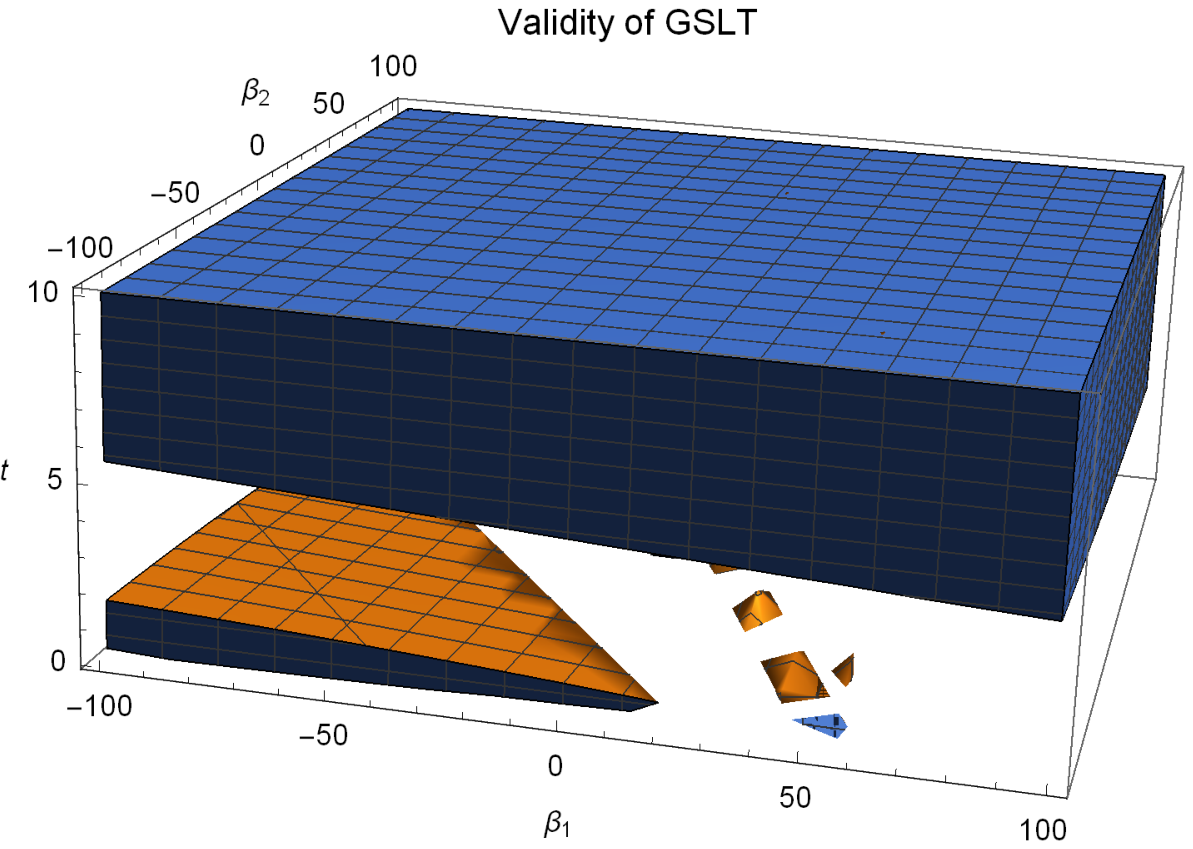,
width=0.5\linewidth} \caption{Left plot represents the validity
regions for logarithmic corrected entropy in intermediate case with
$\beta_3=0$, right graph corresponds to evolution of GST for
$\beta_3=-2$.}\label{fig11}
\end{figure}

\section{Equilibrium Picture}

Here we will talk about the equilibrium picture of first and
generalized second law of thermodynamics. Here we consider the field
equations as
\begin{eqnarray}\nonumber
&&6F_TH^2+(24H^2F_{X_1}+F_{X_2})(3H\dot{H}+\ddot{H})H+F_{X_2}\dot{H}^2+(3H^2-\dot{H})H\dot{F}_{X_2}\\\label{1*n}
&&+24H^3\dot{H}\dot{F}_{X_1}+H^2\ddot{F}_{X_2}+\frac{F}{12}=\rho_m,\\\nonumber
&&2(F_T\dot{H}+H\dot{F}_T+24H[2H\ddot{H}+3(\dot{H}+H^2)\dot{H}]\dot{F}_{X_1}+12H\dot{H}\dot{F}_{X_2}
+24H^2\dot{H}\ddot{F}_{X_1}\\\nonumber
&&+(\dot{H}+3H^2)\ddot{F}_{X_2}+24H^2F_{X_1}\dddot{H}+H\dddot{F}_{X_2}+24F_{X_1}\dot{H}^2(12H^2+\dot{H})\\\label{2*n}
&&+24HF_{X_1}(4\dot{H}+3H^2)\ddot{H})=-p_m,
\end{eqnarray}
It is worthwhile to mention here that all the discussion about first
law of thermodynamics as presented in section \textbf{III} is same
except there is no entropy production term. It is due to the fact
that there is no effective coupling term $\kappa_{eff}$ defined in
the FRW field equations and consequently, the usual energy
conservation equation for both ordinary matter as well as energy and
pressure due to torsion scalar remain satisfied. Also, one need to
use the usual form of Bekenstein-Hawking entropy relationship given
by $S=\frac{A}{4G}$ instead of its modified version. Thus, in this
case, the first law of thermodynamics takes the form
\begin{equation}\nonumber
T_Ad\tilde{S}_A=-dE+dW
\end{equation}
The relationship between entropy due to matter and energy sources
inside the horizon $S_{in}$ and the density and pressure in the
horizon as provided by Gibb's equation can be written as
\begin{equation}\label{3*n}
T_{in}d\tilde{S}_{in}=d(\rho_mV)+p_{m}dV
\end{equation}
which can also be expressed as
\begin{equation}\label{4*n}
T_{in}\dot{\tilde{S}}_{in}=4\pi{\tilde{r}^2_A}(\rho_m+p_m)(\dot{\tilde{r}}_A-H\tilde{r}_A).
\end{equation}
Using the relations for $\rho_m$ and $p_m$ computed from
Eqs.(\ref{1*n}) and (\ref{2*n}), we get the GSLT in the following
form
\begin{eqnarray}\nonumber
T_{A}\dot{\tilde{S}}_{tot}&=&-\frac{\dot{H}}{2GH^4}(2H^2+\dot{H})-\frac{2\pi\kappa^2}{H^4}
(H^2+\dot{H})[6F_TH^2+6(24H^2F_{X_1}\\\nonumber
&+&F_{X_2})(3H\dot{H}+\ddot{H})H+6F_{X_2}\dot{H}^2+18H^3\dot{F}_{X_2}-6H\dot{H}\dot{F}_{X_2}+144H^3\dot{H}\dot{F}_{X_1}\\\nonumber
&+&6H^2\ddot{F}_{X_2}+\frac{F}{2}-2F_T\dot{H}-2H\dot{F}_T-48H\{2H\ddot{H}+3(\dot{H}+H^2)\dot{H}\}
\dot{F}_{X_1}\\\nonumber
&-&24H\dot{H}\dot{F}_{X_2}-48H^2\dot{H}\ddot{F}_{X_1}-2(\dot{H}+3H^2)\ddot{F}_{X_2}-48H^2F_{X_1}\dddot{H}-2H\dddot{F}_{X_2}\\\label{5*n}
&-&48F_{X_1}\dot{H}^2(12H^2+\dot{H})-48HF_{X_1}(4\dot{H}+3H^2)\ddot{H}]\geq0.
\end{eqnarray}
The derivative terms present in this constraint will be evaluated by
using chain rule as follows:
\begin{eqnarray}\label{6*n}
\dot{F}_{X_1}&=&F_{X_1T}\dot{T}+F_{X_1X_1}\dot{X}_1+F_{X_1X_2}\dot{X}_2,\\\label{7*n}
\dot{F}_{X_2}&=&F_{X_2T}\dot{T}+F_{X_2X_1}\dot{X}_1+F_{X_2X_2}\dot{X}_2,\\\nonumber
\ddot{F}_{X_1}&=&F_{X_1TT}\dot{T}^2+2F_{X_1X_1T}\dot{X}_1\dot{T}+2F_{TX_1X_2}\dot{T}\dot{X_2}
+F_{X_1T}\ddot{T}+F_{X_1X_1X_1}\dot{X}_1^2\\\label{9*n}
&+&2F_{X_1X_1X_2}\dot{X}_1\dot{X}_2+F_{X_1X_1}\ddot{X}_1+F_{X_1X_2X_2}\dot{X}_2^2+F_{X_1X_2}\ddot{X}_2,\\\nonumber
\ddot{F}_{X_2}&=&F_{X_2TT}\dot{T}^2+2F_{X_2X_1T}\dot{X}_1\dot{T}+2F_{TX_2X_2}\dot{T}\dot{X_2}
+F_{X_2T}\ddot{T}+F_{X_2X_1X_1}\dot{X}_1^2\\\label{9*n}
&+&2F_{X_2X_1X_2}\dot{X}_1\dot{X}_2+F_{X_2X_2}\ddot{X}_2+F_{X_2X_2X_2}\dot{X}_2^2+F_{X_2X_2}\ddot{X}_2,\\\nonumber
\dddot{F}_{X_2}&=&F_{X_2TTT}\dot{T}^3+3F_{X_2TTX_1}\dot{X}_1\dot{T}^2+3F_{X_2X_2TT}\dot{X}_2\dot{T}^2+3\dot{T}\ddot{T}F_{X_2TT}\\\nonumber
&+&3F_{X_2TX_1X_1}\dot{X}_1^2\dot{T}+F_{X_2TX_1}(\ddot{X}_1\dot{T}+\dot{X}_1\ddot{T})+F_{X_2TX_2}(\ddot{X}_2\dot{T}
+\dot{X}_2\ddot{T})\\\nonumber
&+&3F_{X_2X_2X_2T}\dot{X}_2^2\dot{T}+F_{X_2TX_1}\dot{X}_1\ddot{T}+F_{X_2TX_2}\dot{X}_2\ddot{T}+F_{X_2T}\dddot{T}\\\nonumber
&+&3F_{X_2X_1TX_2}\dot{T}\dot{X}_1\dot{X}_2+F_{X_2X_1T}(\ddot{T}\dot{X}_1+\dot{T}\ddot{X}_1)+F_{X_2X_1X_1X_1}\dot{X}_1^3\\\nonumber
&+&F_{X_2X_1X_1X_2}\dot{X}_1^2\dot{X}_2+F_{X_2X_1X_2X_2}\dot{X}_1\dot{X}_2^2+F_{X_2X_1X_2}(\ddot{X}_1\dot{X}_2+\dot{X}_1\ddot{X}_2)\\\nonumber
&+&3F_{X_2X_1X_1}\dot{X}_1\ddot{X}_1+F_{X_2X_1T}\dot{T}\ddot{X}_1+F_{X_2X_1X_2}\ddot{X}_1\dot{X}_2+F_{X_2X_1}\dddot{X}_1\\\nonumber
&+&F_{X_2X_2T}(\ddot{X}_2\dot{T}+\dot{X}_2\ddot{T})+F_{X_2X_1X_1T}\dot{T}\dot{X}_1\dot{X}_2+F_{X_2X_2X_1X_1}\dot{X}_1^2\dot{X}_2\\\nonumber
&+&2F_{X_2X_2X_2X_1}\dot{X}_1\dot{X}_2^2+F_{X_2X_2X_1}(\ddot{X}_1\dot{X}_2+\dot{X}_1\ddot{X}_2)+F_{X_2X_2X_2X_2}\dot{X}_2^3\\\label{10*n}
&+&3F_{X_2X_2X_2}\dot{X}_2\ddot{X}_2+F_{X_2X_2T}\dot{T}\ddot{X_2}+F_{X_2X_2X_1}\dot{X}_1\ddot{X}_2+F_{X_2X_2}\dddot{X}_2,\\\label{11*n}
\dot{F}_T&=&F_{TT}\dot{T}+F_{TX_1}\dot{X}_1+F_{TX_2}\dot{X}_2.
\end{eqnarray}
Here we will investigate the validity of GSLT given by constraint
Eq.(\ref{5*n}) in this generalized teleparallel gravity. For this
purpose, we examine the compatibility of this constraint for two
specific forms of generic function $F(T, X_1, X_2)$ in the upcoming
subsections.

\subsection{The Validity of GSLT for $F$ Independent of $X_2$}

Here we will explore the validity of GSLT using the form of $F$
given by Eq.(\ref{1*}). The GSLT constraint (\ref{5*n}), in this
case, takes the following form
\begin{eqnarray}\nonumber
T_h\dot{\tilde{S}}_{tot}&=&-\frac{\dot{H}}{2GH^4}(2H^2+\dot{H})-\frac{2\pi\kappa^2}{H^4}(H^2+\dot{H})\times[(2\dot{H}-6H^2)\\\nonumber
&\times&\{\frac{2\alpha_1X_1}{T^3}+\frac{4\delta\alpha_2X_1}{T^5}e^{\frac{\delta
X_1}{T^4}}\}+(144H^3(3H\dot{H}+\ddot{H})-48H\ddot{H}(4\dot{H}+3H^2)\\\nonumber
&-&48H^2\dddot{H}-48\dot{H}^2(12H^2+\dot{H}))
\{\frac{\alpha_1}{T^2}+\frac{\alpha_2\delta}{T^4}e^{\frac{\delta
X_1}{T^4}}\}+(144H^3\dot{H}-48H\\\nonumber
&\times&(2H\ddot{H}+3\dot{H}(\dot{H}+H^2)))
\{(-\frac{4\alpha_2\delta^2X_1}{T^9}-\frac{4\alpha_2\delta}{T^5})\dot{T}+(\frac{\alpha_2\delta^2}{T^8}e^{\frac{\delta
X_1}{T^4}})\dot{X}_1\}\\\nonumber
&+&\frac{1}{2}\left(T+\frac{\alpha_1X_1}{T^2}+\alpha_2e^{\frac{\delta
X_1}{T^4}}\right)-2H\{(\frac{6\alpha_1X_1}{T^4}+\frac{20\delta
X_1\alpha_2}{T^6}e^{\frac{\delta
X_1}{T^4}}+\frac{16\alpha_2\delta^2X_1^2}{T^{10}}e^{\frac{\delta
X_1}{T^4}})\dot{T}\\\nonumber
&-&(\frac{4\alpha_2\delta^2X_1}{T^9}+\frac{4\alpha_2\delta}{T^5})e^{\frac{\delta
X_1}{T^4}}\dot{X}_1\}-48H^2\dot{H}
\{(\frac{36\alpha_2\delta^2X_1}{T^{10}}+\frac{20\alpha_2\delta}{T^6}+\frac{16\alpha_2\delta^3X_1^2}{T^{14}}\\\nonumber
&+&\frac{16\alpha_2\delta^2X_1}{T^{10}})e^{\frac{\delta
X_1}{T^4}}\dot{T}^2+2(-\frac{4\alpha_2\delta^2}{T^9}
+\frac{\delta}{T^4}\left(\frac{4\alpha_2\delta^2X_1}{T^9}+\frac{4\alpha_2\delta}{T^5}\right))e^{\frac{\delta
X_1}{T^4}}\dot{X}_1\dot{T}\\\nonumber
&-&(\frac{4\alpha_2\delta^2X_1}{T^9}+\frac{4\alpha_2\delta}{T^5})e^{\frac{\delta
X_1}{T^4}}\ddot{T} +(\frac{\alpha_2\delta^3}{T^{12}}e^{\frac{\delta
X_1}{T^4}})\dot{X}_1^2 +(\frac{\alpha_2\delta^2}{T^8}e^{\frac{\delta
X_1}{T^4}})\ddot{X}_1\}]\geq0.\\\label{2**}
\end{eqnarray}
Here we will investigate this GSLT condition validity for previously
defined four different forms of expansion factor.

For the de Sitter case, it is seen that the validity of GSLT can be
achieved for some particular choice of involved free parameter
$\alpha_i$ as given in Table \textbf{1}. Specifically, the GSLT
condition is satisfied if $\alpha_2=3.11$. For the case of
cosmographic parameters, some relevant useful derivative terms
appearing in the constraint (\ref{2**}) can be expressed as
\begin{eqnarray}\nonumber
\dot{X}_1\dot{T}&=&-3456H^{10}(1+q)^2\{1+q)^2+j+3q+2\},\quad
\ddot{T}=-12H^4\{(1+q)^2+(j+3q+2)\},\\\nonumber
\dddot{T}&=&12H^5\{3(1+q)(j+3q+2)-(s-2j-5q-3)\},\\\label{4**}
\ddot{X}_1&=&288H^8\{(1+q)^4+5(1+q)^2(j+3q+2)+(j+3q+2)^2-(1+q)(s-2j-5q-3)\}.
\end{eqnarray}
Using the recent values of cosmographic parameters, the higher-order
time rates of the terms $T,~ X_1$ and $X_2$ attain the following
values
\begin{eqnarray}\nonumber
\ddot{H}=0.4072, \quad \dddot{H}=-0.5927, \quad \ddddot{H}=16.1570,
\quad \ddot{T}=-3.9214, \quad \dddot{T}=7.8266 \quad
\ddot{X}_1=55.7851.
\end{eqnarray}
By making the use of these values, the possible restrictions on free
model parameters are represented in Figure \textbf{12} and listed in
Table \textbf{I}.
\begin{figure}[H]
\epsfig{file=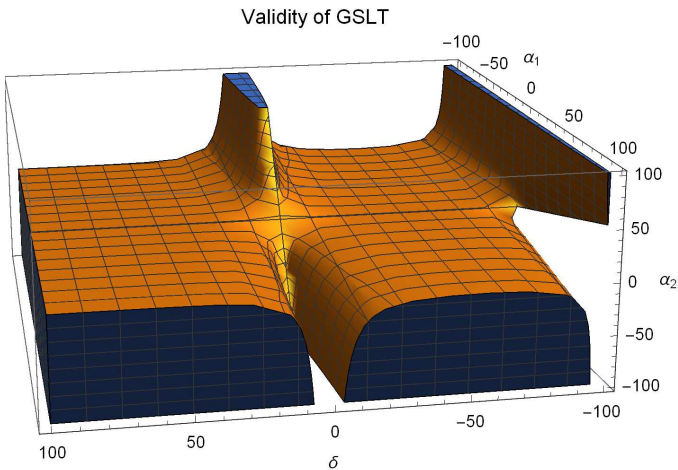,width=0.5\linewidth}\caption{Plot represents
the validity regions of GSLT (\ref{2**}) for equilibrium case in
terms of cosmographic parameters.}\label{fig12}
\end{figure}

Next we consider the possibility of power law form of expansion
factor. The corresponding Hubble parameter, Torsion scalar and
$X_1,~X_2$ terms along with their derivative terms are given by
(\ref{i2}). Some other required higher order time rates can be
calculated as
\begin{eqnarray}\nonumber
\ddot{H}&=& \frac{2b}{(t_s-t)^3}, \quad
\dddot{H}=\frac{6b}{(t_s-t)^4}, \quad
H^{(iv)}=\frac{24b}{(t_s-t)^5},\quad
H^{(v)}=\frac{120b}{(t_s-t)^6},\\\label{5**}
\ddot{X}_1&=&\frac{6048b^4}{(t_s-t)^8},\quad
\dot{X}_1\dot{T}=-\frac{10,368b^6}{(t_s-t)^{10}},\quad
\ddot{T}=-\frac{36b^2}{(t_s-t)^4}.
\end{eqnarray}
For the graphical analysis, the validity of GSLT constraint
(\ref{2**}) is provided by the right part of Figure \textbf{13}.

\begin{figure}[H]
\epsfig{file=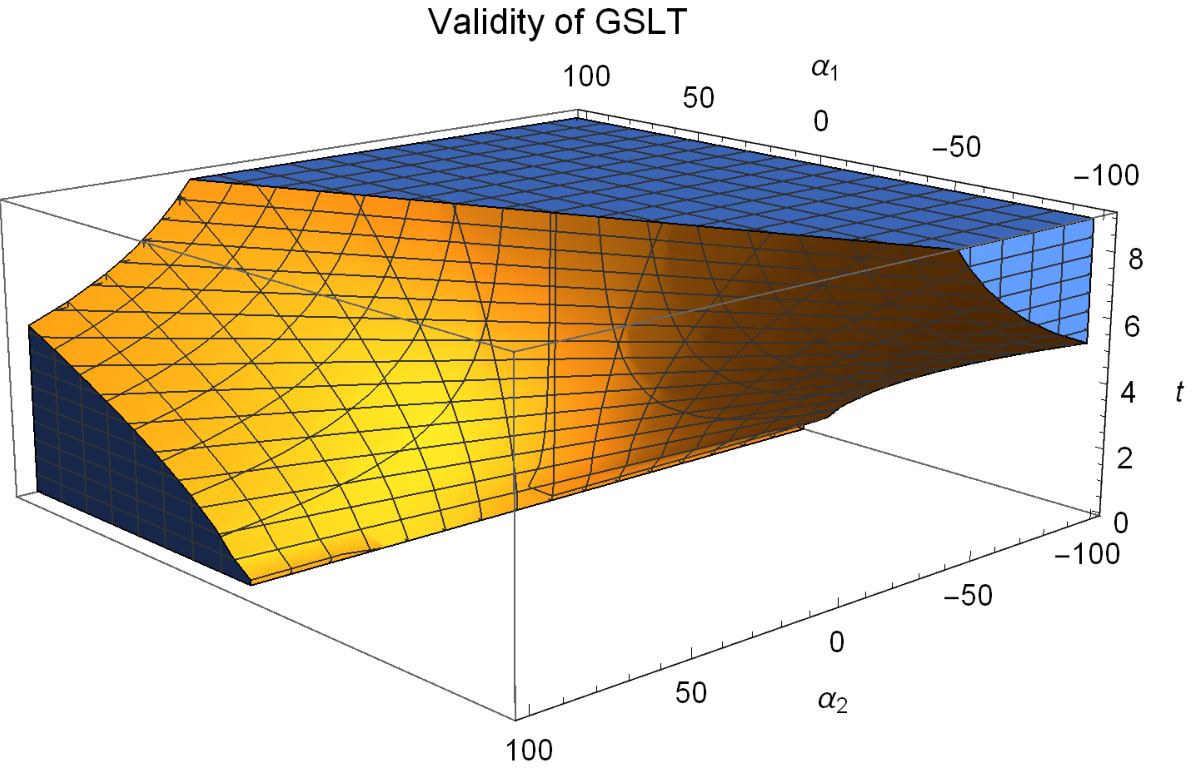,width=0.5\linewidth}\epsfig{file=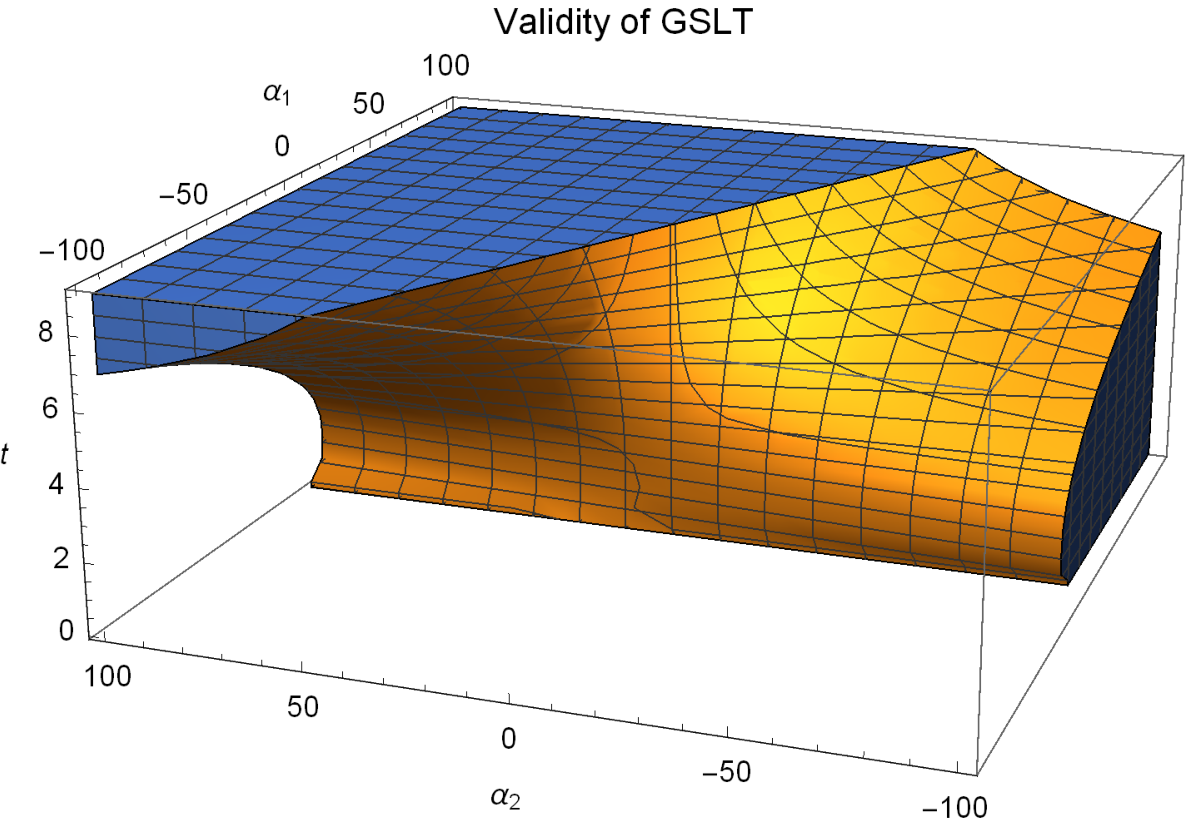,
width=0.5\linewidth} \caption{Left plot represents the validity
regions for logarithmic corrected entropy in intermediate case with
$\delta=2$, right graph corresponds to evolution of GST for
$\delta=-2$.}\label{fig13}
\end{figure}

In last, we will discuss the validity of the GSLT constraint using
the intermediate form of expansion radius. For this form of
expansion radius, the cosmological parameters like Hubble parameter,
torsion scalar, terms $X_1,~X_2$ and its first order time rates are
given by (\ref{i3}). Some other higher order time derivatives
required for the evaluation of GSLT constraint are
\begin{eqnarray}\nonumber
\ddot{H}&=&b_1\beta(\beta-1)(\beta-2)t^{\beta-3},\quad
\dddot{H}=b_1\beta(\beta-1)(\beta-2)(\beta-3)t^{\beta-4},\\\nonumber
H^{(iv)}&=&b_1\beta(\beta-1)(\beta-2)(\beta-3)(\beta-4)t^{\beta-5},\quad
H^{(v)}=b_1\beta(\beta-1)(\beta-2)(\beta-3)(\beta-4)(\beta-5)t^{\beta-6},\\\nonumber
\ddot{T}&=&-12b_1^2\beta^2(\beta-1)(2\beta-3)t^{2\beta-4},\quad
\ddot{X}_1=288b_1^4\beta^4(\beta-1)^2(2\beta-3)(4\beta-7)t^{4\beta-8},\\\label{6**}
\dot{X}_1\dot{T}&=&-3456b_1^6\beta^6(\beta-1)^3(2\beta-3)t^{6\beta-10}.
\end{eqnarray}
In the similar fashion, we find the validity regions for
intermediate cases in the framework of equilibrium picture. It is
mentioned that for the case of $\delta>0$, GSLT is satisfied in
later times in intermediate case. We also present some validity
regions in Figure \ref{fig13},

\subsection{The Validity of GSLT for $F$ Independent of $X_1$}

Here we will consider the form of generic function of $F(T, X_1,
X_2)$ given by Eq.(\ref{1}). Removing the terms depending on $X_1$
and using the above defined form of $F$, the constraint for validity
of GSLT is found as
\begin{eqnarray}\nonumber
T_h\dot{\tilde{S}}_{tot}&=&-\frac{\dot{H}}{2GH^4}\left(2H^2+\dot{H}\right)-\frac{2\pi\kappa^2}{H^4}(H^2+\dot{H})
[(6H^2-2\dot{H})\\\nonumber &\times&\{1-\frac{\beta
X_2}{T^2}-\frac{3\beta_2X_2^2}{T^4}-\frac{3\sigma
X_2\beta_3}{T^4}e^{\frac{\sigma
X_2}{T^3}}\}+(6H(3H\dot{H}+\ddot{H})+6\dot{H}^2)\\\nonumber
&\times&\{\frac{\beta_1}{T}+\frac{2\beta
X_2}{T^3}+\frac{\beta_3\sigma}{T^3}e^{\frac{\sigma
X_2}{T^3}}\}+(18H^3-30H\dot{H})\{(-\frac{\beta_1}{T^2}-\frac{6\beta_2X_2}{T^4}-e^{\frac{\sigma
X_2}{T^3}}\\\nonumber
&\times&(\frac{3\beta_3\sigma^2X_2}{T^7}+\frac{3\beta_3\sigma}{T^4}))\dot{T}
+(\frac{2\beta_2}{T^3}+\frac{\beta_3\sigma^2}{T^6}e^{\frac{\sigma
X_2}{T^3}})\dot{X}_2\}-2\dot{H}\{(\frac{2\beta_1}{T^3}+\frac{24\beta_2X_2}{T^5}\\\nonumber
&+&e^{\frac{\sigma
X_2}{T^3}}\left(\frac{21\beta_3\sigma^2X_2}{T^8}+\frac{12\beta_3\sigma}{T^5}\right)+e^{\frac{\sigma
X_2}{T^3}}(\frac{3\sigma
X_2}{T^4})(\frac{3\beta_3\sigma^2X_2}{T^7}+\frac{3\beta_3\sigma}{T^4}))\dot{T}^2]\\\nonumber
&+&2\dot{X}_2\dot{T}(-\frac{6\beta_2}{T^4}-e^{\frac{\sigma
X_2}{T^3}}\frac{3\beta_3\sigma^2}{T^7}-e^{\frac{\sigma
X_2}{T^3}}(\frac{3\beta_3\sigma^3X_2}{T^{10}}+\frac{3\beta_3\sigma^2}{T^7}))
+(-\frac{\beta_1}{T^2}\\\nonumber
&-&\frac{6\beta_2X_2}{T^4}-e^{\frac{\sigma
X_2}{T^3}}(\frac{3\beta_3\sigma^2X_2}{T^7}+\frac{3\beta_3\sigma}{T^4}))\ddot{T}
+(\frac{\beta_3\sigma^3}{T^9}e^{\frac{\sigma
X_2}{T^3}})\dot{X}_2^2+(\frac{2\beta_2}{T^3}\\\nonumber
&+&\frac{\beta_3\sigma^2}{T^6}e^{\frac{\sigma
X_2}{T^3}})\ddot{X}_2\}+\frac{1}{2}\left(T+\frac{\beta_1
X_2}{T}+\frac{\beta_2X_2^2}{T^3}+\beta_3e^{\frac{\sigma
X_2}{T^3}}\right)-2H\{(\frac{2\beta_1X_2}{T^3}\\\nonumber
&+&\frac{12\beta_2X_2^2}{T^5}+\frac{9\sigma^2X_2^2\beta_3}{T^8}e^{\frac{\sigma
X_2}{T^3}}+\frac{12\sigma X_2\beta_3}{T^5}e^{\frac{\sigma
X_2}{T^3}})\dot{T}+(-\frac{\beta_1}{T^2}-\frac{6\beta_2X_2}{T^4}\\\nonumber
&-&e^{\frac{\sigma
X_2}{T^3}}\left(\frac{3\beta_3\sigma^2X_2}{T^7}+\frac{3\beta_3\sigma}{T^4}\right))\dot{X}_2\}
-2H\{(-\frac{6\beta_1}{T^4}-\frac{120\beta_2X_2}{T^6}-e^{\frac{\sigma
X_2}{T^3}}\\\nonumber
&\times&(\frac{168\beta_3\sigma^2X_2}{T^9}+\frac{60\beta_3\sigma}{T^6})-\frac{3\sigma
X_2}{T^4}e^{\frac{\sigma
X_2}{T^3}}(\frac{21\beta_3\sigma^2X_2}{T^8}+\frac{12\beta_3\sigma}{T^5})+e^{\frac{\sigma
X_2}{T^3}}\\\nonumber &\times&\left(\frac{3\sigma
X_2}{T^4}\right)^2(\frac{3\beta_3\sigma^2X_2}{T^7}+\frac{3\beta_3\sigma}{T^4})-e^{\frac{\sigma
X_2}{T^3}}(3\sigma
X_2)(\frac{33\beta_3\sigma^2X_2}{T^{12}}+\frac{24\beta_3\sigma}{T^9}))\dot{T}^3\\\nonumber
&+&3\dot{X}_2\dot{T}^2(-\frac{3\sigma^3\beta_3}{T^{10}}e^{\frac{\sigma
X_2}{T^3}}-\frac{\sigma}{T^3}(\frac{6\beta_3\sigma}{T^7}+\frac{3\sigma^3\beta_3X_2}{T^{10}})e^{\frac{\sigma
X_2}{T^3}})+3\dot{T}\ddot{T}(\frac{2\beta_1}{T^3}\\\nonumber
&+&\frac{24\beta_2X_2}{T^5}+e^{\frac{\sigma
X_2}{T^3}}(\frac{21\beta_3\sigma^2X_2}{T^8}+\frac{12\beta_3\sigma}{T^5})+e^{\frac{\sigma
X_2}{T^3}}\left(\frac{3\sigma
X_2}{T^4}\right)(\frac{3\beta_3\sigma^2X_2}{T^7}\\\nonumber
&+&\frac{3\beta_3\sigma}{T^4}))-3\dot{X}_2^2\dot{T}(\frac{3\sigma^3\beta_3}{T^{10}}e^{\frac{\sigma
X_2}{T^3}}+(\frac{6\beta_3\sigma}{T^7}+\frac{3\sigma^3\beta_3X_2}{T^{10}})\frac{\sigma}{T^3}e^{\frac{\sigma
X_2}{T^3}})\\\nonumber
&+&3\left(\ddot{X}_2\dot{T}+\dot{X}_2\ddot{T}\right)(-\frac{6\beta_2}{T^4}-e^{\frac{\sigma
X_2}{T^3}}(\frac{3\beta_3\sigma^2}{T^7})-e^{\frac{\sigma
X_2}{T^3}}\frac{\sigma}{T^3}(\frac{3\beta_3\sigma^2X_2}{T^7}+\frac{3\beta_3\sigma}{T^4}))\\\nonumber
&+&(-\frac{\beta_1}{T^2}-\frac{6\beta_2X_2}{T^4}-e^{\frac{\sigma
X_2}{T^3}}(\frac{3\beta_3\sigma^2X_2}{T^7}+\frac{3\beta_3\sigma}{T^4}))\dddot{T})
+(\frac{2\beta_2}{T^3}+\frac{\beta_3\sigma^2}{T^6}e^{\frac{\sigma
X_2}{T^3}})\dddot{X}_2\\\label{8***}&+&3(\frac{\beta_3\sigma^3}{T^9}e^{\frac{\sigma
X_2}{T^3}})\dot{X}_2\ddot{X}_2+\frac{\beta_3\sigma^4}{T^{12}}e^{\frac{\sigma
X_2}{T^3}}\dot{X}_2^3\}]\geq0.
\end{eqnarray}
Using the same four choices of expansion radius, we will check the
compatibility of this constraint graphically. It is seen that for de
Sitter model, the GSLT constraint will be satisfied if we fix
$\beta_3=-3.11$. For the case of cosmographic parameters, some
useful higher order derivatives of the term $X_2$ are given by
\begin{eqnarray}\nonumber
\ddot{X}_2&=&-12H^6\{3(j+3q+2)^2-4(1+q)(s-2j-5q-3)+3(s-2j\\\nonumber
&-&5q-3)-18(1+q)(j+3q+2)-6(1+q)^3+(l-5s+10(q+2)j\\\nonumber
&+&30q(q+2)+24)\},\\\nonumber
\dddot{X}_2&=&-12H^7\{10(j+3q+2)(s-2j-5q-3)-5(1+q)(l-5s\\\nonumber
&+&10(q+2)j+30q(q+2)+24)+3(l-5s+10(q+2)j+30q(q+2)\\\label{8**}
&+&24)-24(1+q)(s-2j-5q-3)+36(1+q)^2(j+3q+2)+18(j+3q+2)^2\}
\end{eqnarray}
which turn out to be as $\ddot{X}_2=-17.2684$ and
$\dddot{X}_2=-50.7413$, for recent fix values of cosmographic
quantities. In this case, the possible validity region for the GSLT
constraint is given by Figure \textbf{14} and the detail is provided
in Table \textbf{I}.

Also, for the power law form of scale factor, the higher-order time
rates of $X_2$ turn out to be
\begin{eqnarray}\nonumber
\ddot{X}_2&=&-\frac{720b^2(1+b)}{(t_s-t)^6}, \quad
\dddot{X}_2=-\frac{4320b^2(1+b)}{(t_s-t)^7},\\\nonumber
\dot{X}_2\ddot{X}_2&=&\frac{103680b^4(1+b)^2}{(t_s-t)^{11}}, \quad
\ddot{X}_2\dot{T}=\frac{8640b^4(1+b)}{(t_s-t)^{9}},\\\nonumber
\dot{X}_2\ddot{T}&=&\frac{5184b^4(1+b)}{(t_s-t)^{9}}\quad
\dot{X}_2\dot{T}=\frac{1728b^4(1+b)^2}{(t_s-t)^{8}},\\\nonumber
\dot{T}\ddot{T}&=&\frac{432b^4}{(t_s-t)^{4}}, \quad
\dot{X}_2^2\dot{T}=-\frac{248832b^6(1+b)^2}{(t_s-t)^{13}},\\\label{9**}
\dot{X}_2\dot{T}^2&=&-\frac{20736b^4(1+b)}{(t_s-t)^{11}}.
\end{eqnarray}
Furthermore, for intermediate form of expansion factor, these
derivatives are computed as follows
\begin{eqnarray}\nonumber
\ddot{X}_2&=&-12\{b_1^2\beta^2(\beta-1)(2\beta-3)(2\beta-4)(2\beta-5)t^{2\beta-6}
+3b_1^3\beta^3(\beta-1)\\\nonumber&\times&
(3\beta-4)(3\beta-5)t^{3\beta-6}\},\\\nonumber
\dddot{X}_2&=&-12\{b_1^2\beta^2(\beta-1)(2\beta-3)(2\beta-4)(2\beta-5)(2\beta-6)t^{2\beta-7}
+3b_1^3\beta^3\\\label{10**}&\times&(\beta-1)(3\beta-4)(3\beta-5)(3\beta-6)t^{3\beta-7}\}.
\end{eqnarray}

Introducing these derivatives in the GSLT constraint (\ref{8***}),
we check its validity by making graphical analysis as presented in
Figures \ref{fig15}. Here, in the left plot, we show the validity
regions for $\beta_3=0$, while the right plot indicates the regions
for $\beta_3=-2$. In case of $\beta_3\leqslant0$, we can not find
one particular region of validity, in fact, there are very small
regions as shown in this plot.
\begin{figure}[H]
\epsfig{file=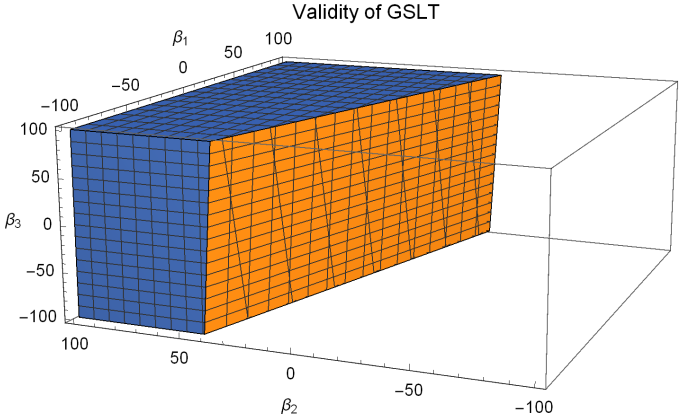, width=0.5\linewidth} \caption{Plot
represents the validity regions for GSLT condition for equilibrium
case in terms of cosmographic quantities for the model
(\ref{1}).}\label{fig14}
\end{figure}
\begin{figure}[H]
\epsfig{file=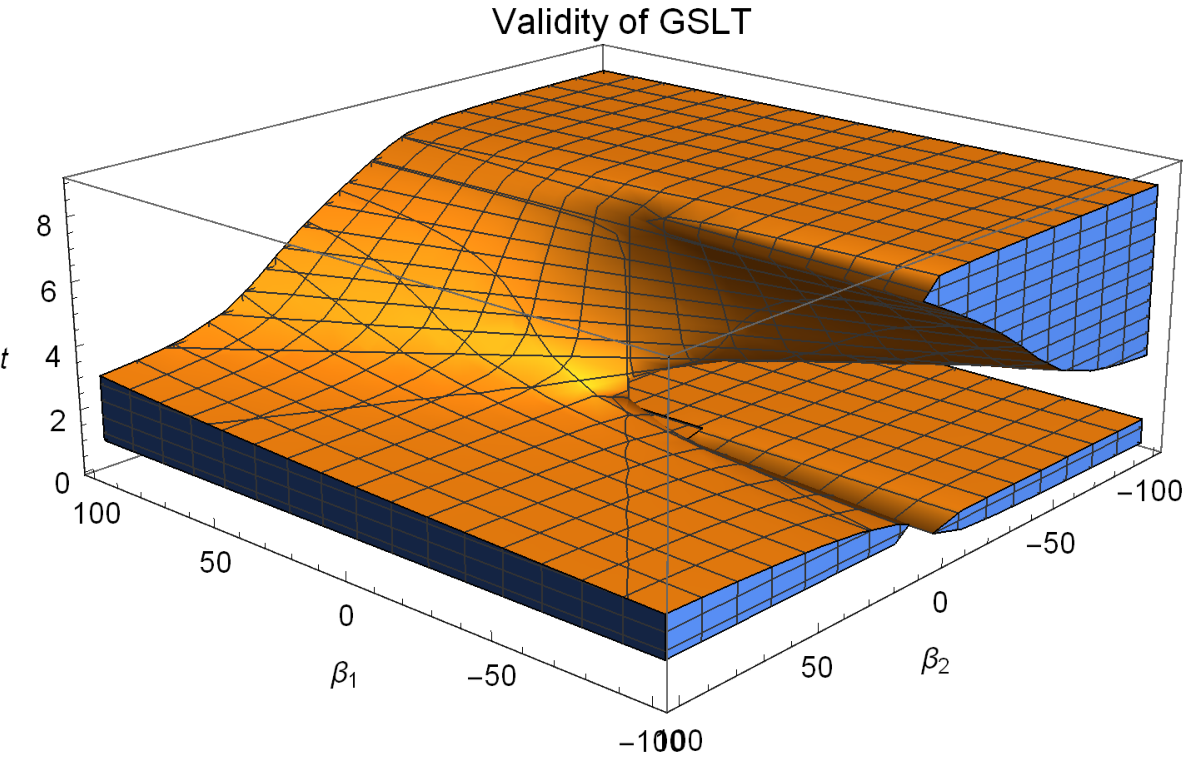,width=0.5\linewidth}\epsfig{file=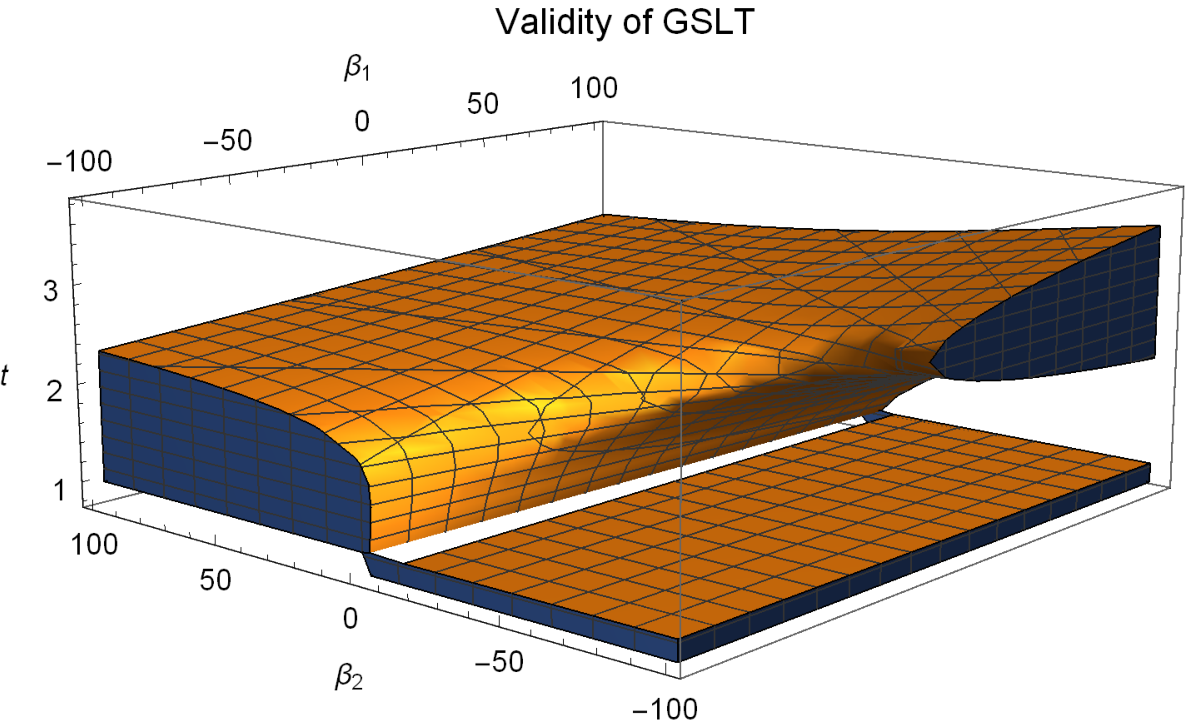,
width=0.5\linewidth} \caption{Left plot represents the validity
regions for equilibrium picture in intermediate case with
$\beta_3=0$, right graph corresponds to evolution of GST for
$\beta_3=-2$.}\label{fig15}
\end{figure}

\section{Concluding Remarks}

\begin{table}
    \centering
    \scriptsize{
\begin{tabular}{|c|c|c|}
\hline
\cline{3-3} $F(T,X_1,X_2)$ Models & Validity of GSLT & Cosmographic Parameters \\
\cline{3-3}
& & $q,~j,~r,~s,~l$ parameters \\
\hline
 &  &
if $\delta\geq20$ \& $\alpha_2\leq-20$, $\forall$ $\alpha_1$  \\
\cline{3-3} Model $1$ & Non-Equilibrium Picture
 & if $\delta\leq0$ \& $\alpha_2\geq0$ with $\alpha_1\leq0$ \\
\cline{2-3}
 & Logarithmic & if $\delta\geq15$ \& $\alpha_2\leq-30$, $\forall$ $\alpha_1$ \\
\cline{3-3}
$F(T,~X_1,~X_2)=$ & Corrected Entropy  & if $\delta\leq-20$ \&
$\alpha_2\geq85$, $\forall$ $\alpha_1$ \\
\cline{2-3}
$T+\frac{\alpha_1X_1}{T^2}+\alpha_2e^{\frac{\delta
X_1}{T^4}}$ & Equilibrium Picture & if $\delta\geq40$ \& $\alpha_2\leq-10$, $\forall$ $\alpha_1$ \\
\cline{3-3}
 &  & if $\delta\leq-15$ \& $\alpha_2\leq-20$, $\forall$ $\alpha_1$ \\
\hline
& Non-Equilibrium Picture &
if $\beta_3\geq70$ $\forall$ $\beta_1$ \& $\beta_2$ \\
\cline{2-3}
Model $2$ & Logarithmic & if $\beta\leq-35$ \& $\beta_3\geq50$ $\forall$ $\beta_2$ \\
$F(T, X_1, X_2)=$ & Corrected Entropy & \\
\cline{2-3}
$T+\frac{\beta_1
X_2}{T}+\frac{\beta_2X_2^2}{T^3}+\beta_3e^{\frac{\sigma X_2}{T^3}}$ & Equilibrium Picture &
if $\beta_2\leq50$, $\forall$ $\beta_1$ \& $\beta_3$\\
\hline
\end{tabular}}
\caption{Validity regions of $\dot{\tilde{S}}_{tot}\geq 0$ for
different models.} \label{Table1}
\end{table}

\begin{table}
    \centering
    \scriptsize{
\begin{tabular}{|c|c|c|c|c|}
\hline
 &  &  \multicolumn{3}{c|}{Various Scale Factors} \\
\cline{3-5}
$F(T,X_1,X_2)$ Models & Validity of GSLT & de-Sitter Model & Power Law Form & Intermediate Form \\
\cline{3-5}
& &$H=H_0$ & $a(t)=a_0(t_s-t)^{-b}$ & $a(t)=e^{b_1 t^\beta}$ \\
\hline
 &  &  & if $\delta>0$; $\alpha_2\geqslant0$ \& $\forall$ $\alpha_1$ &
if $\delta>0$; $\alpha_1\geqslant0$ \& $\alpha_2\leqslant-20$ \\
\cline{4-5} Model $1$ & Non-Equilibrium Picture & Trivially
Satisfied &
 if $\delta=0$, $\alpha_1\geqslant0$ \& $\forall$ $\alpha_2$ &
 if $\delta=0$, $\alpha_1\leqslant0$ \& $\forall$ $\alpha_2$ \\
\cline{4-5}
 & & & if $\delta<0$, $\alpha_1\geqslant0$ \& $\forall$ $\alpha_2$
 & if $\delta<0$; $\alpha_1\geqslant0$ \& $\alpha_2\leqslant-5$ \\
\cline{2-5} $F(T,~X_1,~X_2)=$ &  Logarithmic &
$\lambda_1=-6.06+0.16\lambda_2$, & if $\delta>0$;
$\alpha_2\geqslant10$ \& $\forall$ $\alpha_1$ &
if $\delta>0$; $\alpha_1\leqslant0$ \& $\alpha_2\leqslant-11$  \\
\cline{4-5} $T+\frac{\alpha_1X_1}{T^2}+\alpha_2e^{\frac{\delta
X_1}{T^4}}$ & Corrected Entropy & $\forall$ $\lambda_2$ & if
$\delta=0$; $\alpha_1\leqslant-18$ \& $\forall$ $\alpha_2$ &
if $\delta=0$; $\alpha_1\leqslant-30$ \& $\forall$ $\alpha_2$ \\
\cline{4-5}
 &  & & if $\delta<0$; $\alpha_1\leqslant0$ \& $\forall$ $\alpha_2$
& if $\delta<0$; $(\alpha_1, \alpha_2)\leqslant-15$ \\
\cline{2-5}
 &  &  &
 if $\delta>0$; $\alpha_2\geqslant1$ \& $\forall$ $\alpha_1$
 & if $\delta>0$; $(\alpha_1, \alpha_2)\geqslant0$, Later times \\
\cline{4-5}
 & Equilibrium Picture & $\alpha_2=3.11$ &
 if $\delta=0$; $\alpha_2\leqslant-1$ \& $\forall$ $\alpha_1$
 & if $\delta=0$; $\alpha_1\geqslant0$ \& $\alpha_2\leqslant0$ \\
\cline{4-5}
 &  & &
 if $\delta<0$; $\alpha_2\leqslant-1$ \& $\forall$ $\alpha_1$
 & if $\delta<0$; $\alpha_1\geqslant5$ \& $\alpha_2\leqslant0$ \\
\hline
  & Non-Equilibrium Picture & Trivially Satisfied &
if $\beta_3<0$; $\beta_1$ \& $\beta_2$  &
if $0<\beta_3<80$; $\forall$ $(\beta_1,\beta_2)<0$\\
\cline{5-5} Mode $2$ & & &  &
if $\beta_3\geqslant80$; $\forall$ $\beta_1$ \& $\beta_2$ \\
\cline{4-5}
 &  & & if $\beta_3=0$; $\forall$ $\beta_1$ \& $\beta_2\leqslant-20$ &
if $\beta_3=0$; $\forall$ $(\beta_1,\beta_2)>0$ \\
\cline{2-5}
$F(T, X_1, X_2)=$ &  Logarithmic  & & &  if $\beta_3>0$; $\beta_1\leqslant-80$, $\forall$ $\beta_2$\\
\cline{5-5} $T+\frac{\beta_1
X_2}{T}+\frac{\beta_2X_2^2}{T^3}+\beta_3e^{\frac{\sigma X_2}{T^3}}$
 & Corrected Entropy& Trivially Satisfied &
if $\beta_3=0$; $\beta_1\geqslant0$ \& $\beta_2\leqslant-20$ &
if $\beta_3=0$; $(\beta_1,\beta_2)\geqslant0$\\
\cline{4-5} & & & if $\beta_3<0$; $\forall$ $\beta_1$ \& $\beta_2$ &
if $\beta_3<0$; $\forall$ $\beta_1$ \& $\beta_2$\\
\cline{2-5}
 & Equilibrium Picture & $\beta_3=-3.11$ &
if $\beta_3>0$, $\forall$ $\beta_1$ \& $\beta_2$ &
if $\beta_3>0$, $\forall$ $\beta_1$ \& $\beta_2$ Later times\\
\cline{5-5}
 & &  & &
if $\beta_3<0$, Fig.(\ref{fig10}) \\
\hline
        \end{tabular}}
        \caption{Validity regions of $\dot{\tilde{S}}_{tot}\geq 0$ for different models.} \label{Table2}
        \end{table}
In the present manuscript, we have discussed the laws of
thermodynamics in a generalized gravitational framework based on
higher-order derivatives of torsion scalar. By taking flat FRW model
with barotropic fluid as matter distribution, we have discussed the
FLT and GSLT at Hubble horizon in both equilibrium and
non-equilibrium perspectives. Firstly, we have presented the
non-equilibrium picture of these thermodynamical laws in such
gravity at the Hubble horizon of FRW model. In order to investigate
the validity of resulting inequalities, we have used two specific
models of $F(T, X_1, X_2)$ function and some interesting cases for
scale factor namely, constant Hubble parameter, cosmographic
parameters, power law and intermediate forms. In the same section,
we have explored the validity of GSLT by taking lograithmic
corrected entropy into account. In all cases, we have checked the
validity of GSLT constraints graphically and found the possible
conditions on the involved free model parameters.

In this generalized teleparallel gravity, it is seen that the
gravitational equations can lead to the non-equilibrium picture of
thermodynamical laws due to the presence of an extra entropy
production term based on the function $F(T, X_1, X_2)$. This is
quite similar to the cases of many other modified gravity theories
like where such extra term appeared in FLT of thermodynamics
(\cite{26}, \cite{34}, \cite{35}, \cite{5*} etc.). In this
non-equilibrium picture of thermodynamical laws, we investigated the
validity of GSLT constraint using two models of $F(T, X_1, X_2)$
function both involving inverse and exponential torsion scalar
terms. In the first place, by fixing some of the involved free model
parameters, we explored the ranges of other parameters for which
GSLT constraint remains satisfied using 3D region plots. Then by
taking these interesting ranges of free parameters into account, we
have shown the validity of GSLT graphically in few cases. We have
also investigated the possible ranges of free parameters for the
validity of GSLT constraints in the presence of logarithmic
corrections in entropy relation using region graphs for both $F(T,
X_1, X_2)$ models.

Furthermore, we have investigated the possibility of equilibrium
picture existence of these thermodynamical laws. For previously used
two models of the function $F(T, X_1, X_2)$, we formulated the
resulting GSLT constraints at Hubble horizon and checked their
validity using cosmographic parameters as well as the power and
intermediate forms of expansion radius. A detailed graphical
analysis of these inequalities and the possible restrictions on free
parameters in terms of region graphs have also been presented there.
All the possible constraints on the free parameters in both
equilibrium as well as non-equilibrium perspectives in all cases of
expansion radius can be summarized in the forms of Tables \textbf{I}
and \textbf{II}.

In literature, this higher-order torsion derivatives based theory
has only been investigated for stability analysis using fixed point
theory and the validity of energy condition bounds for restricting
free model parameters. In these discussions, a very limited analysis
of free parameters selection has been provided. However, the present
paper is providing a very detailed analysis of model parameters
selection in order to make them compatible with the GSLT constraint
and hence leading to a positive contribution in the regard. It would
be worthwhile to explore the validity of GSLT constraints at
apparent as well as event horizons in this generalized
teleparallel gravity by constraining the free involved parameters.\\

\end{document}